\documentclass[twocolumn]{aastex631}
\usepackage{bm}
\usepackage{amsmath}
\usepackage[bottom]{footmisc}
\usepackage{hyperref}

\newcommand\Roct{$\mathrm{Ro_c}$}

\newcommand{\rhobar}{\overline{\rho}}
\newcommand{\Tbar}{\overline{T}}
\newcommand{\Pbar}{\overline{P}}

\newcommand{\vnabla}{\bm{\nabla}}
\newcommand{\vvec}{\bm{v}}

\newcommand{\lpeak}{\ell_\mathrm{peak}}

\def\thinazavg{
\begin{figure}[ht!]
\centering
\includegraphics[trim=0 0 0 0,clip,width=0.99\columnwidth]{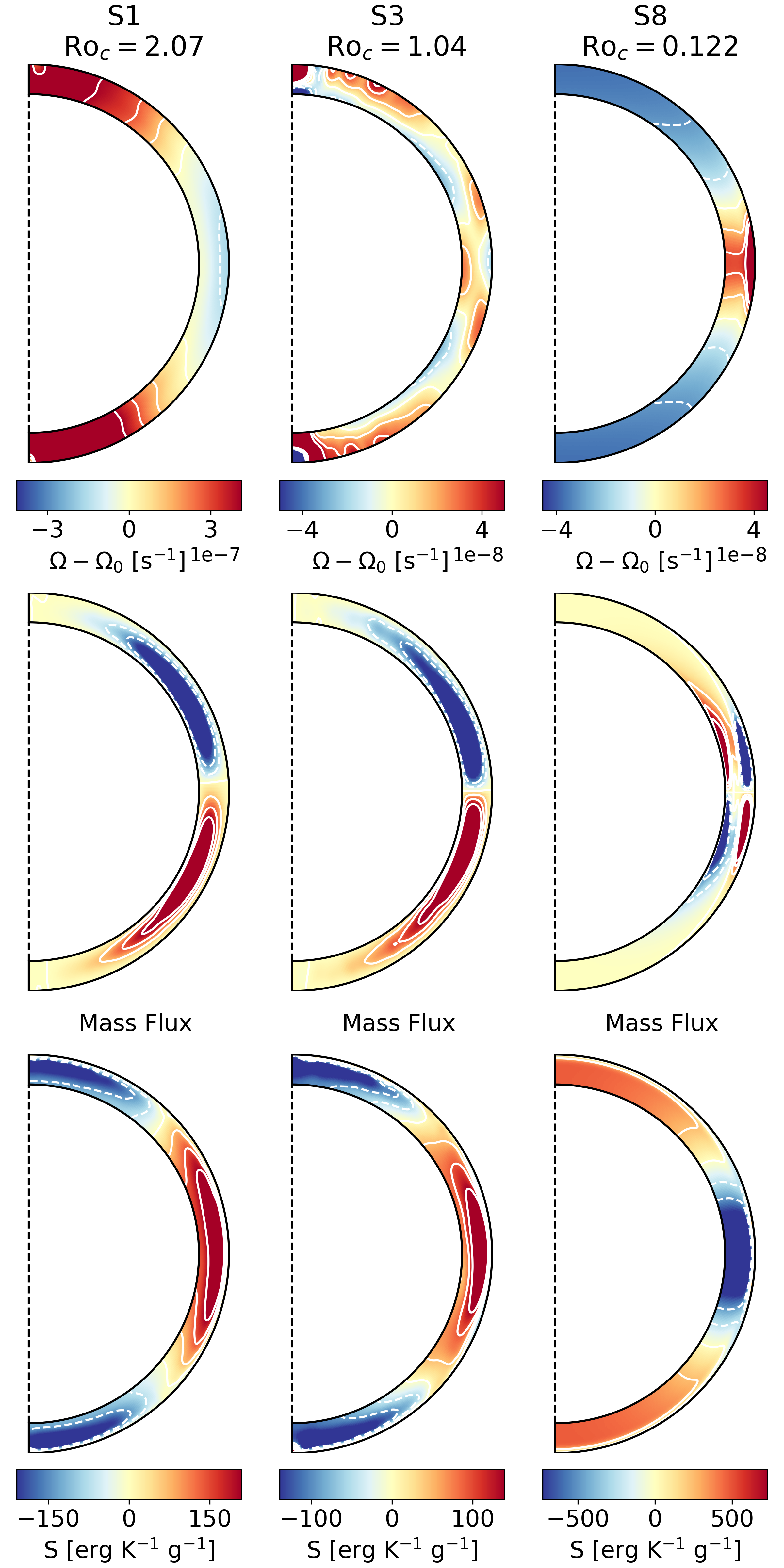}
\caption{Fluid profiles for thin-shell cases in the antisolar (left column;  S1), transitional (central column; S3) and solar-like regimes (right column; S8).  (\textit{Upper Row}):  profiles of differential rotation (angular velocity $\Omega$ in the rotating frame).  (\textit{Central Row}):  Streamlines of meridional mass flux with red (blue) underlay indicating clockwise (counter-clockwise) motion.  (\textit{Lower Row}):  Specific entropy, with spherical mean subtracted.   All profiles have been averaged in time and longitude.  Antisolar cases exhibit rapidly-rotating poles, monocellular meridional circulation within each hemisphere and a warm equatorial region.   Solar-like cases exhibit a rapidly-rotating equator, multicellular meridional flow within a hemisphere and warm poles.   Transitional cases exhibit a combination of solar and antisolar differential rotation, but otherwise possess antisolar characteristics.
}
\label{fig:thin_azavg}
\vspace{-0.1in}
\end{figure}
}

\def\thickazavg{
\begin{figure}[t!]
\centering
\includegraphics[trim=0 0 0 0,clip,width=1.\columnwidth]{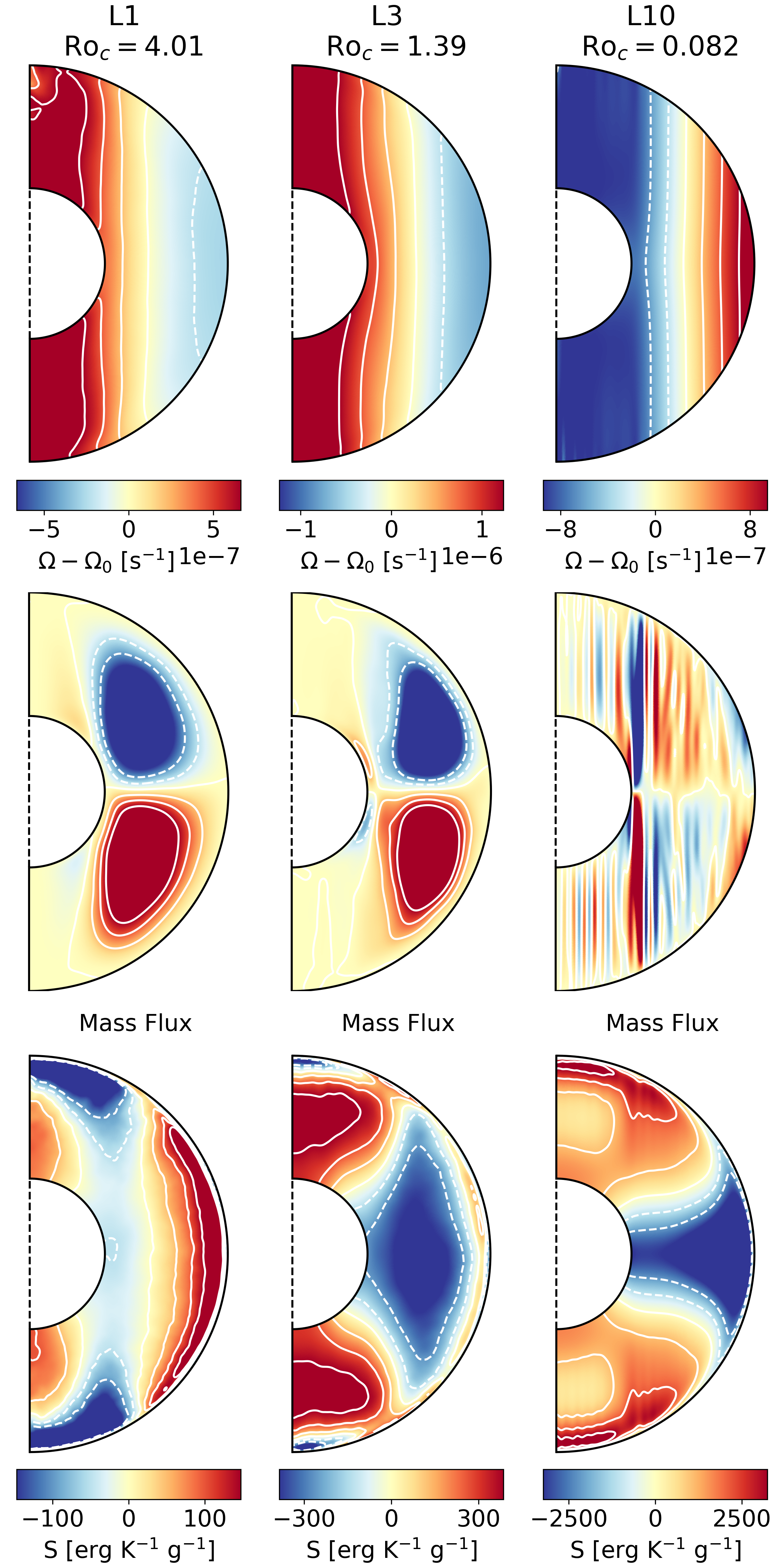}
\caption{Same as Figure \ref{fig:thin_azavg} but depicting a selection of models with thick convection zones (L1, L3 and L10).  Transitional behavior is observed in the specific entropy profile which retains aspects of low-\Roct behavior at depth even as the differential rotation and meridional circulation are antisolar in nature. }
\label{fig:thick_azavg}
\end{figure}
}

\def\thermal{
\begin{figure}[ht!]
\centering
\includegraphics[width=1.\columnwidth]{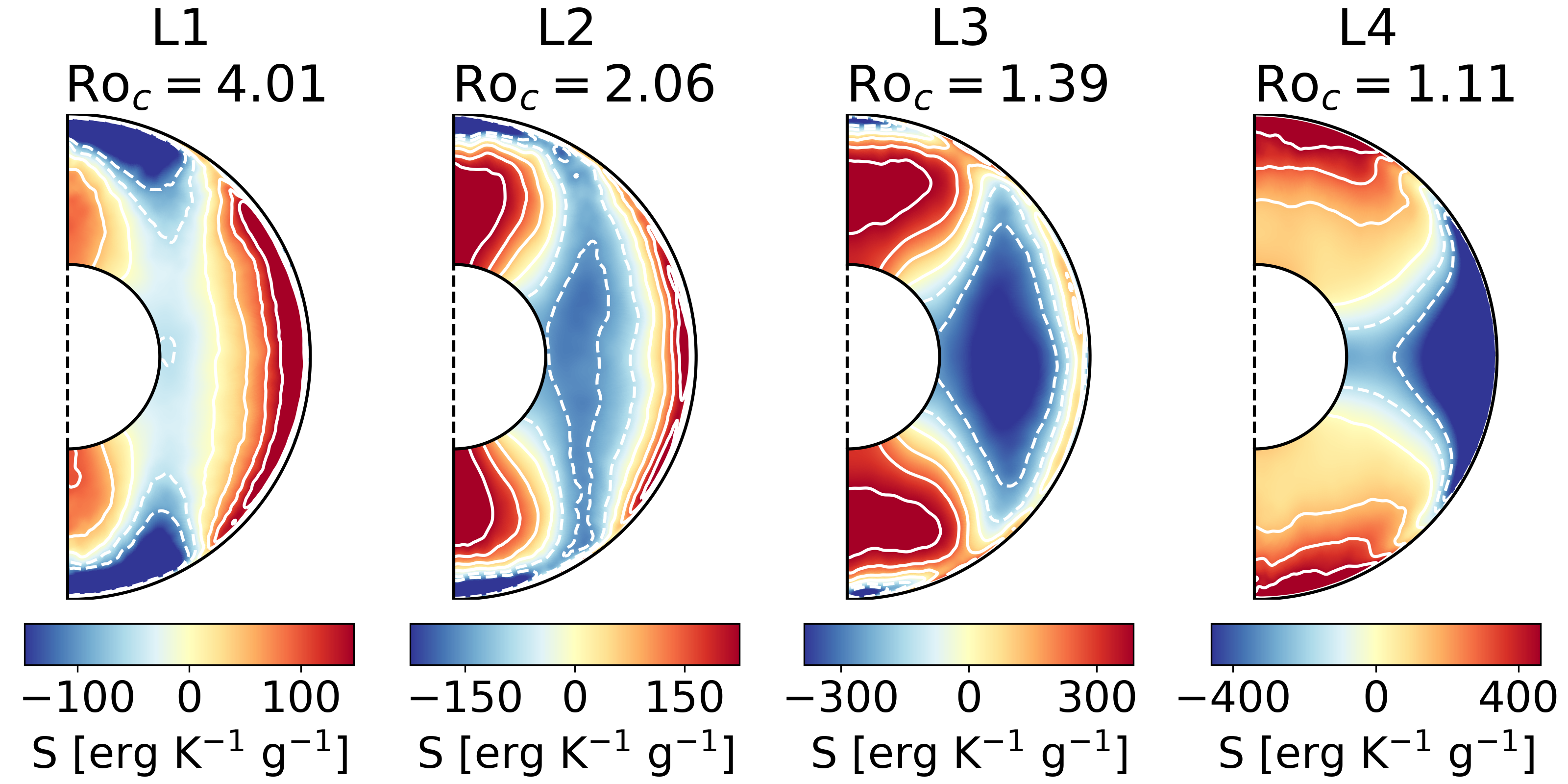}
\caption{Profiles of specific entropy for a series of thick-convection-zone models spanning the solar/antisolar transition.  As in Figures \ref{fig:thin_azavg} and \ref{fig:thick_azavg}, profiles have been averaged in longitude and time, and the spherically symmetric mean has been subtracted.  Case L4 possesses a solar-like differential rotation, where cases L1, L2, and L3 possess antisolar differential rotation.  The transition to an antisolar thermal state is a much broader function of \Roct~than the transition associated with differential rotation in these models.  The cool-pole/warm-equator configuration of the antisolar state manifests first in the upper boundary layer and becomes deeper as \Roct~is increased. }
\label{fig:thermal}
\end{figure}
}

\def\thinflows{
\begin{figure}[t!]
\includegraphics[trim=55 0 55 0,clip,width=\columnwidth]{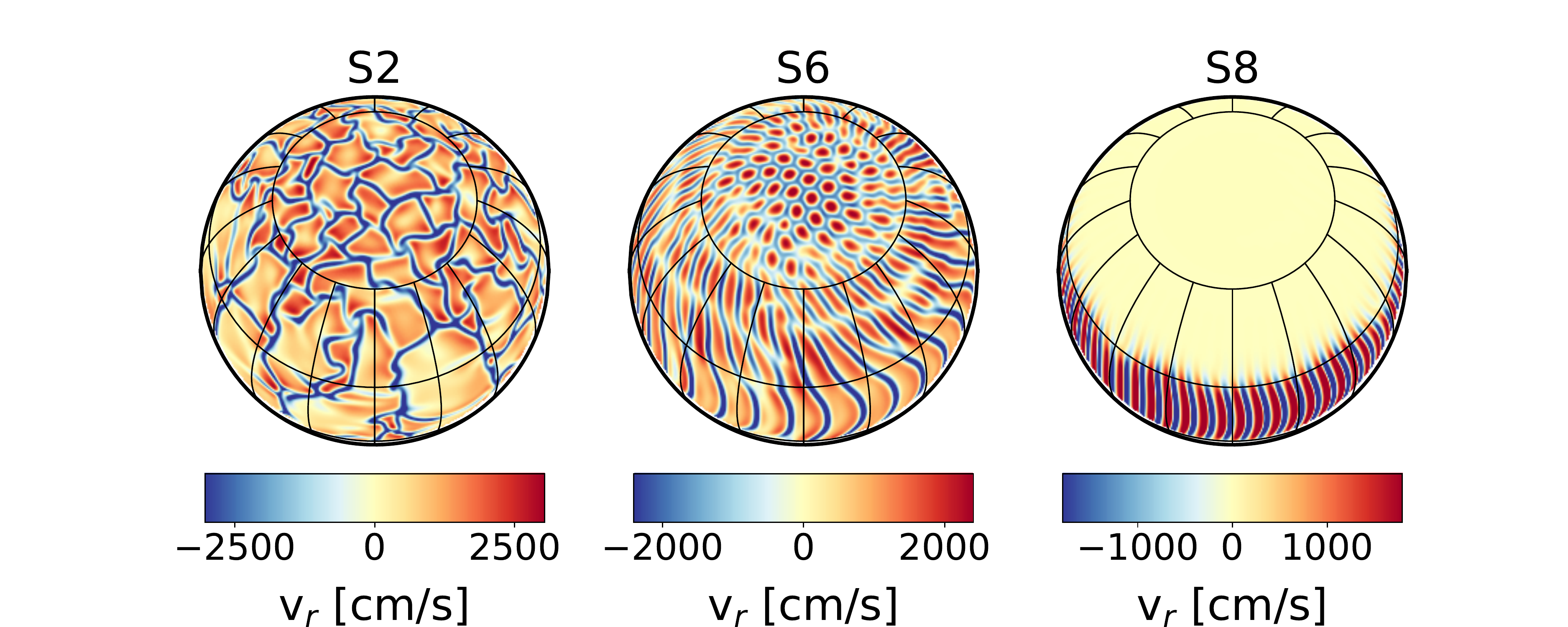}
\includegraphics[trim=10 0 50 30,clip,width=\columnwidth]{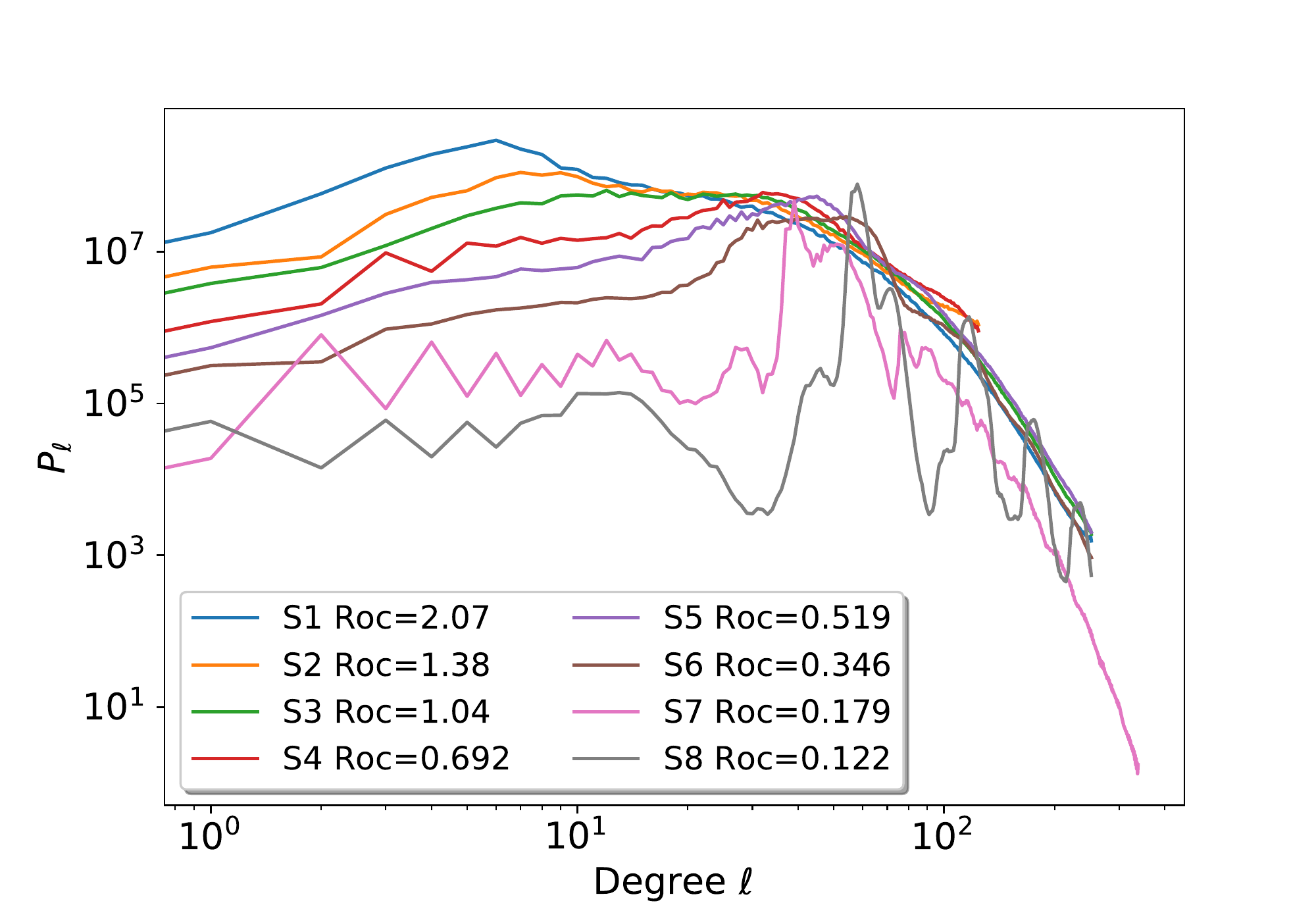}
\caption{Variation of convective structure as a function of convective Rossby number \Roct~for the thin-shell models in this study. (\textit{Upper panels}): Snapshots of radial velocity $v_r$ taken near the outer boundary ($r/r_\mathrm{outer}\sim 0.99$) of three representative models (S2, S6 and S8) with a thin convection zone.  Upflows are colored in red and downflows are colored in blue. (\textit{Lower panel}): Time-averaged spherical harmonic spectra of horizontal convective velocity power $P_\ell$ plotted for all thin-convection zone models, computed at $r/r_\mathrm{outer}\sim 0.99$. As $Ro_c$ decreases, spectral power peaks at higher $\ell$-values and the associated convective structures (upper panel) become thinner in azimuth.   Models S7 and S8 are near convective onset and evince multiple prominent convective peaks as a result.
}
\label{fig:thin}
\end{figure}
}

\def\thickflows{
\begin{figure}[ht!]
\includegraphics[trim=55 0 55 0,clip,width=\columnwidth]{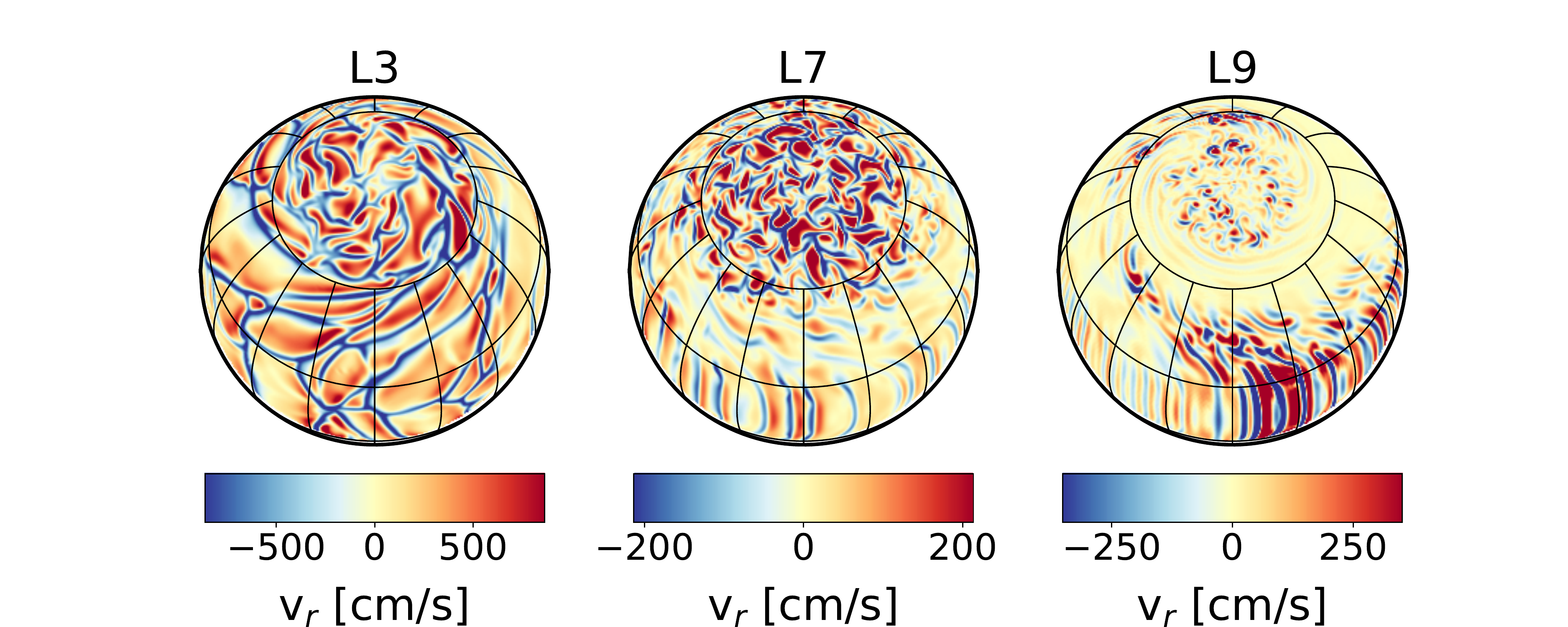}
\includegraphics[trim=10 0 50 30,clip,width=\columnwidth]{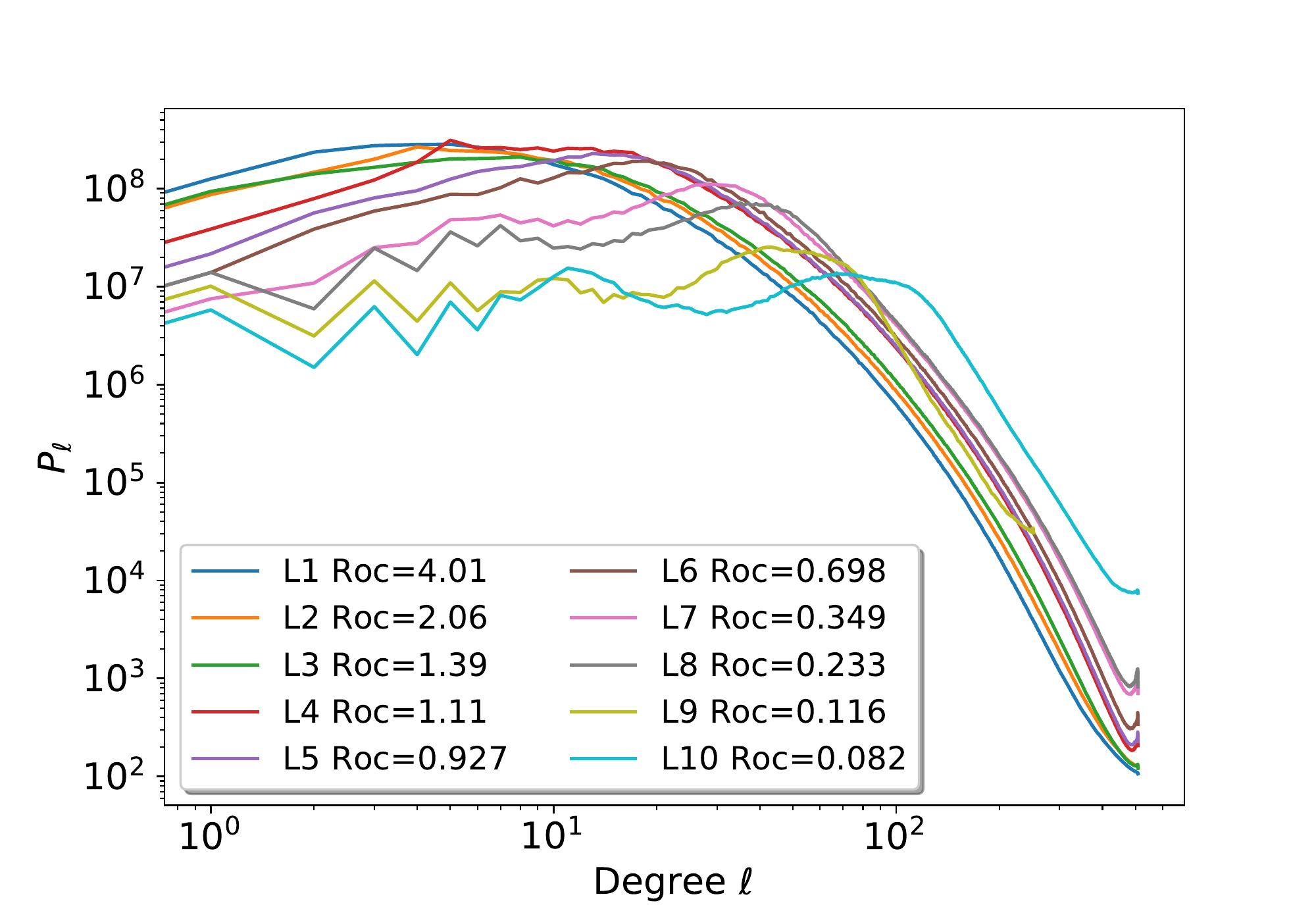}
\caption{Same as Figure \ref{fig:thin}, but for those models with a thick convective zone.  (\textit{Upper row}):  representative radial velocity sampled near the outer boundary ($r/r_\mathrm{outer}\sim 0.99$) for three selected thick-convection zone models.   (\textit{Lower panel}):  Time-averaged convective power $P_\ell$ for all thick-convection-zone models examined in this study. }
\label{fig:thick}
\end{figure}
}

\def\transitionfig{
\begin{figure*}[ht!]
\includegraphics[width=\textwidth]{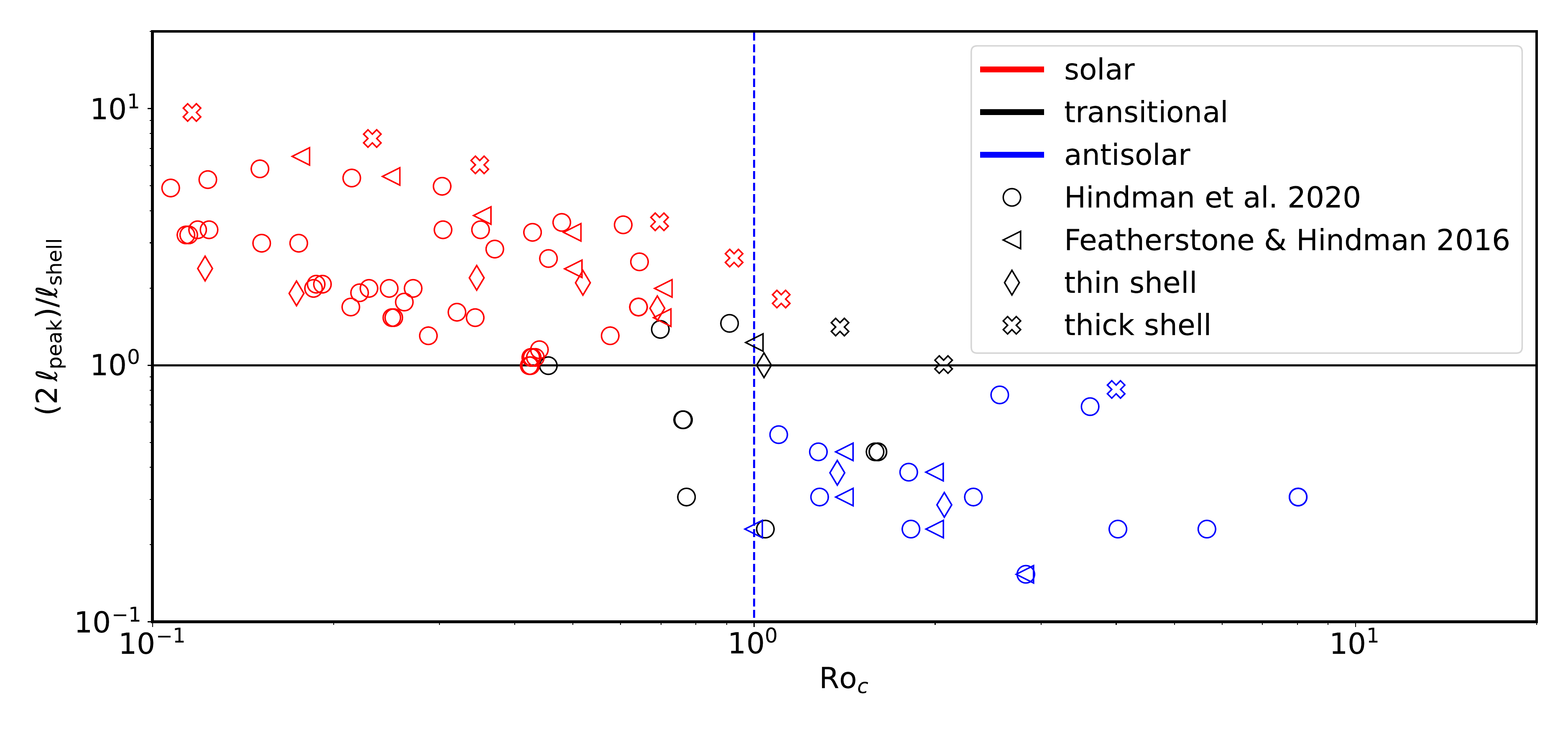}
\caption{ Solar, antisolar, and transitional behavior for all models considered in this study, displayed as a function of characteristic convective wavenumber $\ell_\mathrm{peak}$ and \Roct.  The value of $\ell_\mathrm{peak}$ has been normalized by $\ell_\mathrm{shell}/2$, the wavenumber associated with twice 
the shell depth.  Circles and triangles indicate data previously reported in \citet{Hindman20} and \cite{FH16rot} respectively.  As found in prior studies, \Roct=1 delineates the transition between solar and antisolar behavior.   The transition is equally well-characterized by the point at which the characteristic convective wavelength is equal to twice the shell depth (i.e., when $\ell_\mathrm{peak}$ and $\ell_\mathrm{shell}/2$ are equal.)}
\label{fig:transition}
\end{figure*}
}

\def\composite{
\begin{figure*}
\centering
\includegraphics[trim=190 200 160 20,clip,width=2.\columnwidth]{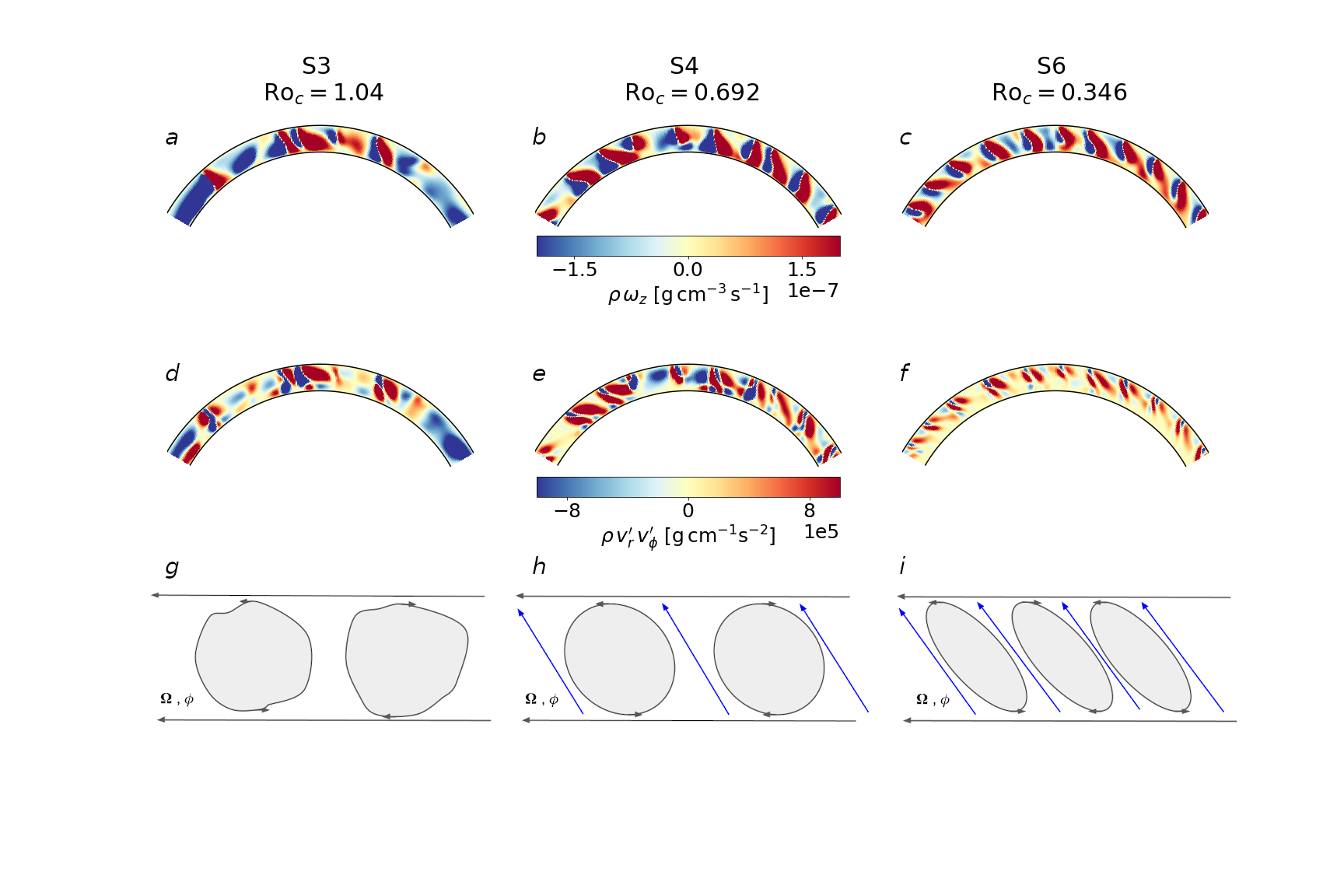}
\caption{Changing convective flow structure as the solar/antisolar transition is approached.  (\textit{Upper row, a--c}):  Snapshots of the $z$-component of vorticity, weighted by density, in the equatorial plane for thin shell cases with three different values of \Roct.  A common color scaling has been adopted for all three models  (\textit{Center row, d--f}):  Companion view of corresponding convective  Reynolds-stress correlation $\rhobar\, v_r' v_\phi' $. As in the upper row, a common color scale has been adopted.   (\textit{Lower row, g--i}): Schematic of convective flow pattern at each \Roct, sketched in the equatorial plane. Blue arrows indicate the direction of angular momentum transport due to convective Reynolds stress in cases S4 and S6.   As \Roct\, approaches unity, columns become wider and the prograde tilting becomes less pronounced.  At \Roct\, of unity, the convective scale is approximately equal to the layer depth and tilting is no longer possible.  The correlations in $v_r'$ and $v_\phi'$ required to transport angular momentum equatorward are lost as a result.
}
\label{fig:composite}
\end{figure*}
}

\def\scaling{
\begin{figure}[t!]
\centering
\includegraphics[trim=20 20 20 0,clip,width=1.\columnwidth]{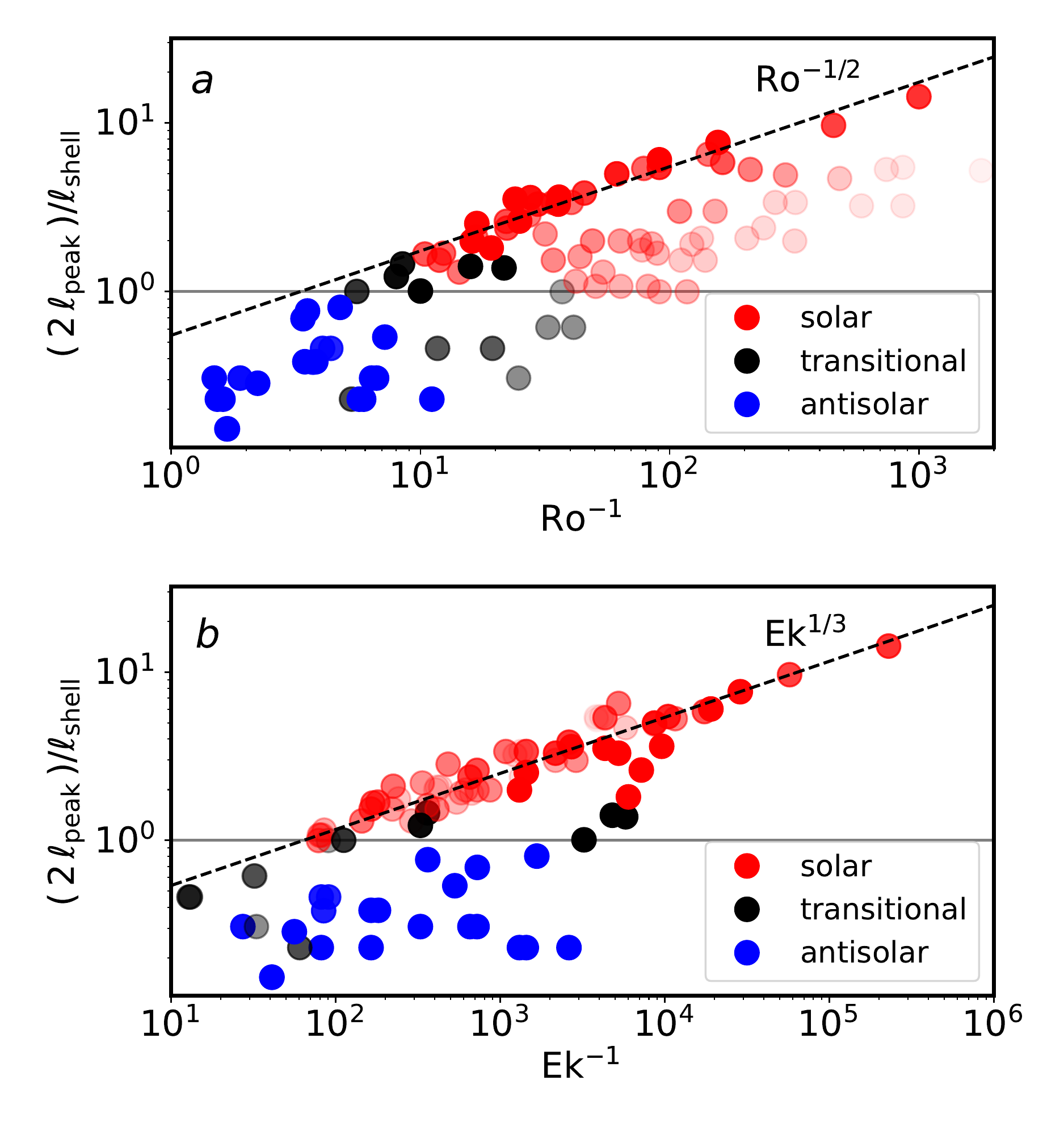}
\caption{  { Scaling of normalized peak convective wavenumber $2\,\ell_\mathrm{peak}/\ell_\mathrm{shell}$ with respect to (\textit{a}) system-scale Rossby number Ro and (\textit{b})  Ekman number Ek for all models considered in this study.  A solid gray reference line at $2\,\ell_\mathrm{peak}/\ell_\mathrm{shell}=1$ has been plotted in each panel.  Mean-flow classification is indicated by symbol color. Symbols are shaded based on the logarithm of reduced Rayleigh number $\mathrm{Ra}^*=\mathrm{Ra}_\mathrm{F}\,\mathrm{Ek}^{4/3}$.  Lower (higher) values of $\mathrm{Ra}^*$ are indicated by lighter (darker) shading.  As $\mathrm{Ra}^*$ increases, models in the solar-like regime tend toward the dashed Ro$^{-1/2}$ reference line associated with the CIA balance.  Regardless of the degree of supercriticality, solar-like models tend to follow the $\mathrm{Ek}^{1/3}$ onset scaling as indicated by the dashed reference line in panel \textit{b}.
}
}
\label{fig:scaling}
\end{figure}
}

\def\FsFz{
\begin{figure}[t!]
\centering
\includegraphics[trim=20 0 10 0,clip,width=1.\columnwidth]{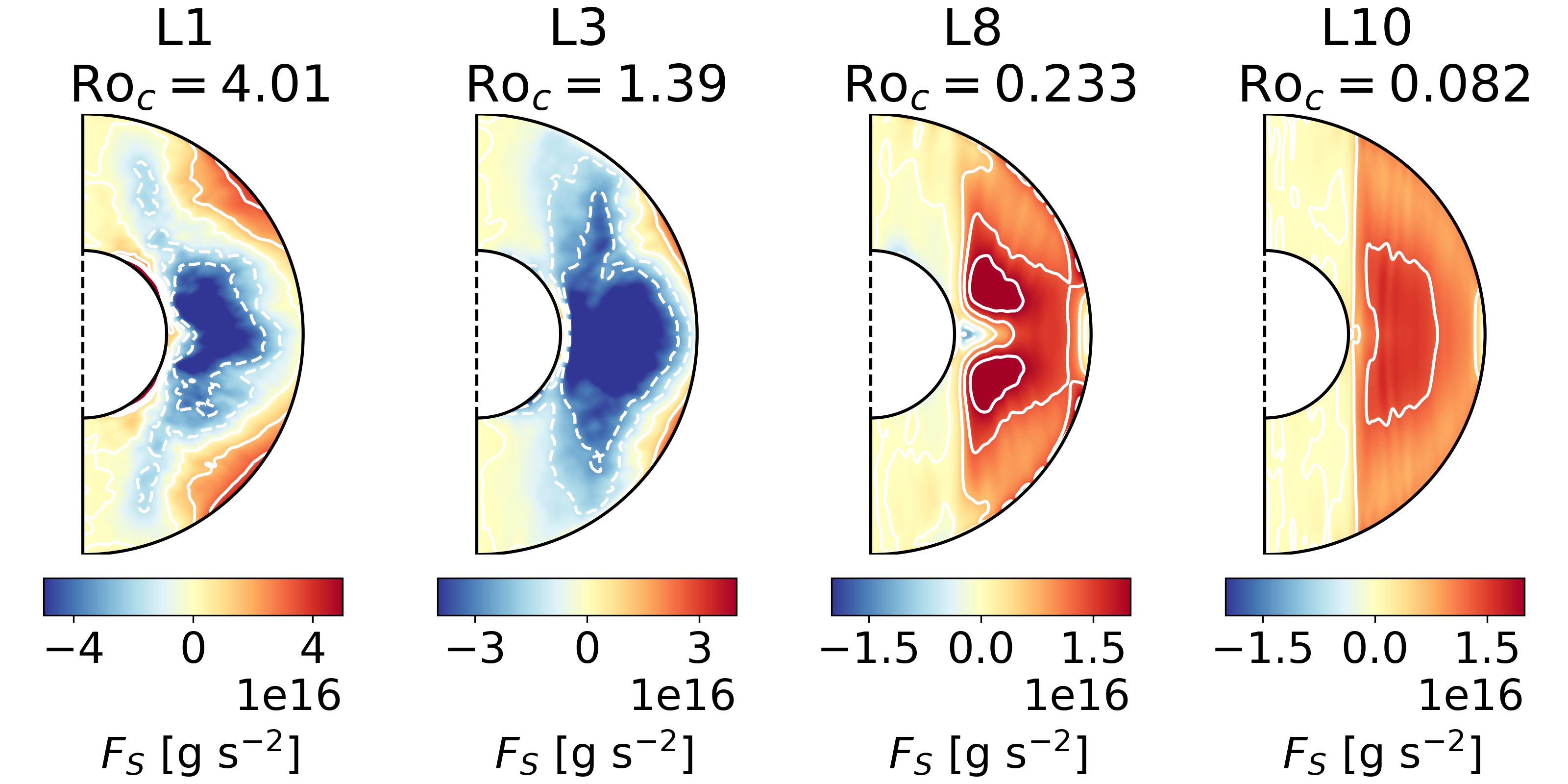}
\includegraphics[trim=20 25 10 0,clip,width=1.\columnwidth]{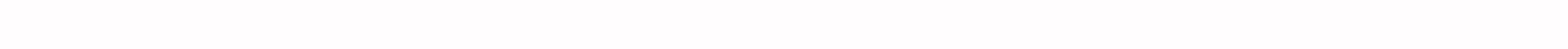}
\includegraphics[trim=20 0 10 55,clip,width=1.\columnwidth]{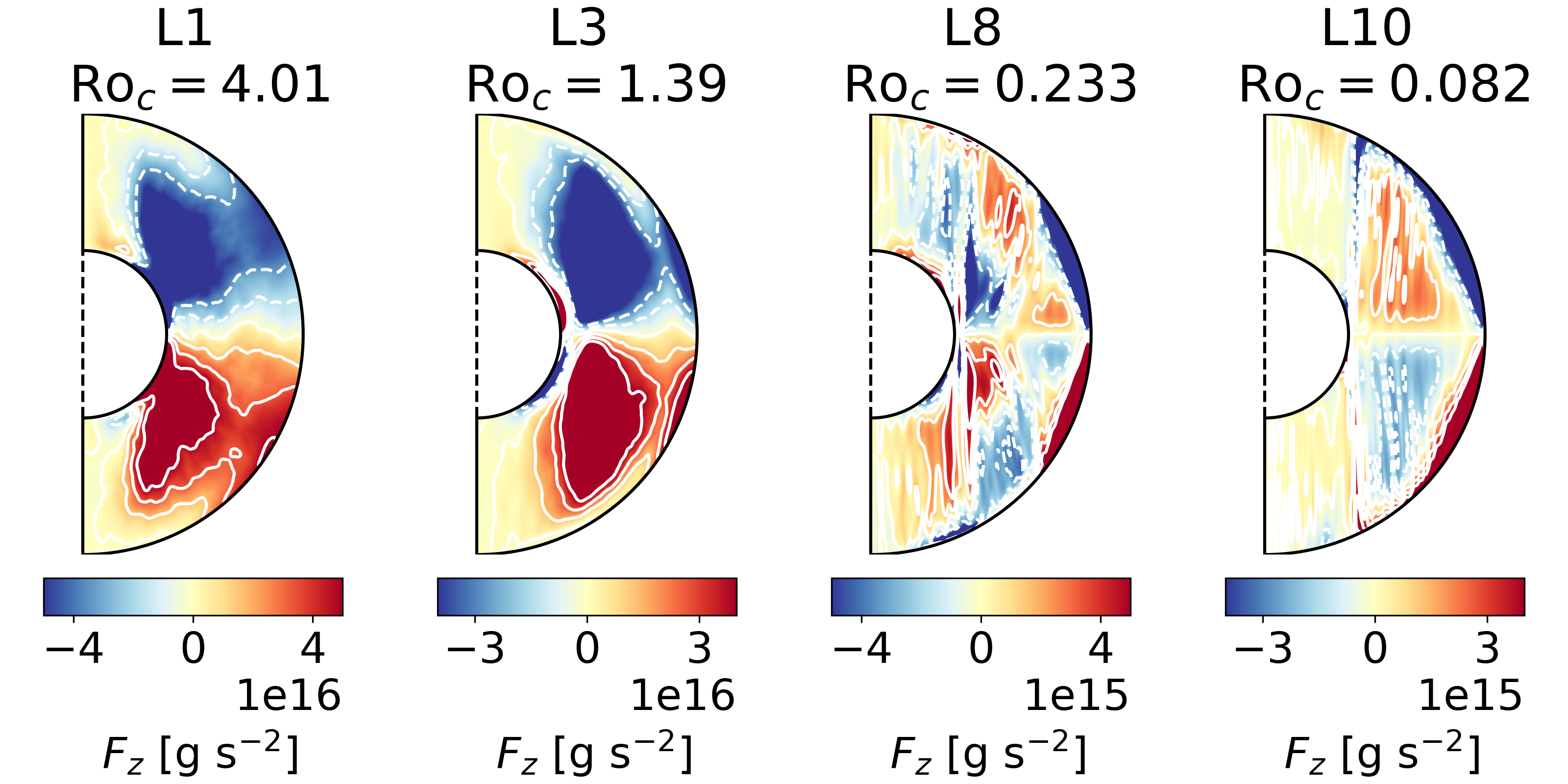}
\caption{{Convective Reynolds stress, decomposed into cylindrical coordinates, as realized in a selection of the thick-shell models.  Profiles have been averaged in longitude and time.  (\textit{upper row}) Component of Reynolds stress in the cylindrical radial direction ($F_s$) for models in the antisolar, transitional and solar-like regimes.  (\textit{lower row})  Corresponding view of the $z$-component ($F_z$).   Low-Ro systems exhibit strong transport of angular momentum away from the rotation axis, but only weak transport in the parallel direction. Transitional and high-Ro$_\mathrm{c}$ systems possess strong transport in both directions, tending to transport angular momentum both inward toward the rotation axis and toward the equatorial plane.  Strong anistropy between the two directions arises only in the low-Ro$_\mathrm{c}$ systems.
}}
\label{fig:FsFz}
\end{figure}
}

\def\thintable{
\begin{deluxetable*}{cccccccccc}
\tablenum{1}
\tablecaption{Physical properties of our models with a thin convective zone. For all models presented, the inner radius $r_{\rm inner}=5.0\times 10^{10}$\,cm, outer radius $r_{\rm outer}=5.88\times 10^{10}$\,cm, shell aspect ratio $\chi=0.85$, and luminosity $L_\star=0.74\, \rm L_\odot$ (where $\rm L_\odot=3.839 \times 10^{33} erg\,s^{-1}$), except for S8, where we considered it to be $1.47\, \rm L_\odot$. Thermal diffusivity and viscosity are considered constant through the convective layer, with values $\kappa=\nu=4\times 10^{12}\, \rm cm^2/s$. Simulation checkpoints, snapshots, and time-averaged outputs for each model are available at {\color{blue} \href{https://osf.io/j275z/wiki/Dataset\%20Listing/} {https://osf.io/j275z/wiki/Dataset\%20Listing/}}  
}
\tablewidth{0pt}

\tablehead{ &\multicolumn6c{Input parameters} & \multicolumn3c{Output parameters}\\
\colhead{Name} & \colhead{Angular velocity $\Omega$} & \colhead{Ek} & \colhead{$Ro_c$} &
\colhead{$Ra_F$} & \colhead{$N_r$} & \colhead{$\ell_{\rm max}$}& \colhead{$Ro$} &
\colhead{$\ell_{\rm peak}$} & \colhead{solar/antisolar DR}}
\startdata
S1 & $1.44\times 10^{-6}\, \rm  s^{-1}$ & $1.78\times 10^{-2}$ & 2.07 & $1.35\times 10^4$ & 128 & 255 & 0.45 & 6 & A\\
S2 & $2.16\times10^{-6}\, \rm  s^{-1}$ & $1.19\times10^{-2}$ & 1.38 & $1.35\times10^4$ & 64 & 127 & 0.27 & 8 & A\\
S3 & $2.87\times10^{-6}\, \rm  s^{-1}$ & $8.95\times10^{-3}$ & 1.04 & $1.35\times10^4$ & 64 & 255 & 0.18 & 21 & T\\
S4 & $4.31\times 10^{-6}\, \rm  s^{-1}$ & $5.96\times 10^{-3}$ & 0.692 & $1.35\times 10^4$ & 64 & 127 & 0.096 & 35 & S\\
S5 & $5.74\times 10^{-6}\, \rm  s^{-1}$ & $4.48\times 10^{-3}$ & 0.519 & $1.35\times 10^4$ & 64 & 255 & 0.06 & 44 & S\\
S6 & $8.61\times 10^{-6}\, \rm  s^{-1}$ & $2.98\times 10^{-3}$ & 0.346 & $1.35\times 10^4$  & 64 & 255 & 0.0316 & 46 & S\\
S7 & $1.72\times 10^{-5}\, \rm  s^{-1}$ & $1.49\times 10^{-3}$ & 0.179 & $1.35\times 10^4$  & 64 & 341 & 0.00815 & 40 & S\\
S8 & $3.44\times 10^{-5}\, \rm  s^{-1}$ & $7.46\times 10^{-4}$ & 0.122 & $2.69\times 10^4$ & 64 & 255 & 0.0042 & 50 & S\\
\enddata
\end{deluxetable*}
}

\def\thicktable{
\begin{deluxetable*}{cccccccccc}
\tablenum{2}
\tablecaption{Physical properties of our models with a thick convective zone. For all models presented, %$Ra_F=4.441\times 10^7$
inner radius $r_{\rm inner}=5.0\times 10^{10}$\,cm, outer radius $r_{\rm outer}=13.16\times 10^{10}$\,cm, shell aspect ratio
$\chi=0.38$ and luminosity $L_\star=4\, \rm L_\odot$. Thermal diffusivity and viscosity are considered constant 
through the convective layer, with values $\kappa=\nu=4\times 10^{12}\,{\rm cm^2/s}$, except for L10, where we considered $\kappa=\nu=2\times 10^{12}\, {\rm cm^2/s}$.  Simulation data for each model is available at {\color{blue} \href{https://osf.io/j275z/wiki/Dataset\%20Listing/} {https://osf.io/j275z/wiki/Dataset\%20Listing/}}  }

\tablewidth{0pt}

\tablehead{ &\multicolumn6c{Input parameters} & \multicolumn3c{Output parameters}\\
\colhead{Name} & \colhead{Angular velocity $\Omega$} & \colhead{Ek}  & \colhead{$Ro_c$} &
\colhead{$Ra_F$} & \colhead{$N_r$} & \colhead{$\ell_{\rm max}$} & \colhead{$Ro$} &
\colhead{$\ell_{\rm peak}$} & \colhead{solar/antisolar DR}}
\startdata
L1 & $0.5\times 10^{-6} \, \rm  s^{-1}$ & $6.01\times 10^{-4}$ & 4.01 & $4.441 \times 10^7$ & 256 & 511 & 0.21 & 4 & A\\
L2 &  $0.97\times 10^{-6} \, \rm  s^{-1}$ & $3.10\times 10^{-4}$ & 2.06 & $4.441 \times 10^7$ & 256 & 511 & 0.10 & 5 & T \\
L3 & $1.44\times 10^{-6} \, \rm  s^{-1}$ & $2.09\times 10^{-4}$ & 1.39 & $4.441 \times 10^7$ & 256 & 511 & 0.063 & 7 & T\\
L4 & $1.8\times 10^{-6} \, \rm  s^{-1}$ & $1.67\times 10^{-4}$ & 1.11 & $4.441 \times 10^7$ & 256 & 511 & 0.052 & 9 & S\\
L5 & $2.16\times 10^{-6}\, \rm  s^{-1}$ & $1.39\times 10^{-4}$ & 0.927 & $4.441 \times 10^7$ & 256 & 511 & 0.04 & 13 & S\\
L6 & $2.87\times 10^{-6}\, \rm  s^{-1}$ & $1.05\times 10^{-4}$ & 0.698 & $4.441 \times 10^7$ & 512 & 511 & 0.0278 & 18 & S\\
L7 & $5.74\times 10^{-6}\, \rm  s^{-1}$ & $5.24\times 10^{-5}$ & 0.349 & $4.441 \times 10^7$ & 512 & 511 & 0.011 & 30 & S\\
L8 & $8.61\times 10^{-6}\, \rm  s^{-1}$ & $3.49\times 10^{-5}$ & 0.233 & $4.441 \times 10^7$ & 256 & 511 & 0.0064 & 38 & S\\
L9 & $1.72\times 10^{-5}\, \rm  s^{-1}$ & $1.75\times 10^{-5}$ & 0.116 & $4.441 \times 10^7$ & 256 & 255 & 0.0022 & 48 & S\\
L10 & $3.44\times 10^{-5}\, \rm  s^{-1}$ & $4.36\times 10^{-6}$ & 0.082 & $3.553 \times 10^8$ & 256 & 511 & 0.001 & 71 & S\\
\enddata
\end{deluxetable*}
}

\shorttitle{Solar-like to Antisolar
Differential Rotation}
\shortauthors{Camisassa \& Featherstone}

\graphicspath{{./}{figures/}}

\begin{document} %https://www.overleaf.com/project/60c2a3bfd8560a2499512929

\title{Solar-like to Antisolar
Differential Rotation:
A Geometric Interpretation}

\author[0000-0002-3524-190X]{Maria E. Camisassa}
\affiliation{Department of Applied Mathematics, University of Colorado, Boulder, CO 80309-0526, USA}
\email{maria.camisassa@colorado.edu}

\author[0000-0002-2256-5884]{Nicholas A. Featherstone}
\affiliation{Southwest Research Institute, Department of the Space Studies, Boulder, CO 80302, USA}

\begin{abstract}

%250 word limit 

The solar convection zone rotates differentially, with its equatorial region rotating more rapidly than the polar regions. This form of differential rotation, also observed in many other low-mass stars, is understood to arise when Coriolis effects are stronger than those associated with buoyant driving of the convection. When buoyancy dominates, a so-called antisolar state of differential rotation results, characterized by rapidly-rotating poles and a slow equator. The transition between these two states has been shown to occur when the intensity of these two forces is roughly equal or, equivalently, when the convective Rossby number of the system is unity. Here we consider an alternative view of the transition that relates this phenomenon to convective structure and convective-zone depth.  Using a series of 3-D rotating convection-zone simulations, we demonstrate that the solar/antisolar transition occurs when the columnar convective structures characteristic of rotating convection attain a diameter roughly equivalent to the shell depth.  When the characteristic convective wavelength exceeds twice the shell depth, we find that the coherent convective structures necessary to sustain an equatorward Reynolds stress are lost, and an antisolar state results.  We conclude by presenting a force-balance analysis that relates this geometric interpretation of the transition to the convective-Rossby-number criteria identified in previous studies.

\end{abstract}


\section{Introduction} \label{sec:intro}

Since the 17th century, observations of sunspots in the Sun have demonstrated that its surface rotates differentially \citep{1655iedi.book.....G}. The rotation period of the poles is roughly 30 days, compared to the fast-rotating equator that completes one revolution in 24 days.  Moreover, the internal rotation profile of the Sun has been shown by helioseismology to exhibit significant latitudinal gradients of shear throughout the convection zone, as well as layers of strong radial shear at its base and in the near-photospheric regions  \citep{2003ARA&A..41..599T,Howe2009}.

Advances in stellar observational techniques have subsequently enabled the measurement of differential rotation profiles in other stars.  Those measurements have revealed that most stars exhibit a solar-like differential rotation, characterized by a fast equator and slow poles \citep{2002MNRAS.330..699C, 2005MNRAS.357L...1B,2006MNRAS.370..468M,2011MNRAS.413.1939M,2006A&A...446..267R,2008MNRAS.390..545D,2011MNRAS.411.1301J,2018Sci...361.1231B,2019A&A...623A.125B}.   Observations have also revealed another rotational regime in a few giants, subgiants and dwarfs.  This regime, characterized by a slowly-rotating equator and rapidly-rotating polar regions, is known as ``antisolar'' differential rotation \citep{2003A&A...408.1103S,2005AN....326..287W,2015A&A...573A..98K,2017A&A...606A..42K,2016A&A...592A.117H,2018Sci...361.1231B}.

While observational confirmation of antisolar behavior in stars has been a relatively recent development, computational studies of rotating convection have long posited the existence of such a state \citep[e.g.,][]{Gilman1977}.  Within the solar context, the solar/antisolar transition is of particular interest due to its link to the Sun's large-scale meridional circulation which mediates the timing of the solar cycle in many dynamo models \citep[e.g.,][]{Dikpati1999,Charbonneau2020}.  While the meridional flow is well-characterized in the upper convective zone, helioseismic observations yield conflicting descriptions of its deep structure.  Some studies indicate (whereas others do not) the presence of multiple cells in depth \citep[][]{Schad2012,Zhao2013,Jackiewicz2015,Gizon2020}.  

Numerical studies of differential rotation may help to clarify this ambiguity owing to the fact that meridional flow is driven in response to convective angular momentum transport via a process known as gyroscopic pumping \citep{Miesch2011,Featherstone2015}.  Systems evincing solar-like differential rotation tend to possess multiple meridional cells in depth, whereas antisolar states tend to possess monocellular flow within each hemisphere \citep{Gastine2013,2014MNRAS.438L..76G,Guerrero2013,Featherstone2015}.  Only if the Sun was in a transitional state, a possibility with some observational support \citep{Metcalfe2016}, would a monocellular meridional flow be expected to occur along with a rapidly-rotating equator.

The transition between these two regimes of differential rotation has also been explored through studies of zonal winds in giant planets \citep{2007Icar..190..110A,Gastine2013,Soderlund2013}.  In fact, while Jupiter and Saturn possess a complex, banded-wind structure, the ice giants Uranus and Neptune posses relatively simple surface-rotation profiles that are antisolar in nature \citep{Sukoriansky2002,Helled2010}.  Whether motivated by planetary or stellar considerations, a large body of work suggests that the transition between solar and antisolar states is controlled by the Rossby number, Ro, of the convecting fluid.  This nondimensional number expresses the ratio of rotational and convective timescales.  Specifically,
\begin{equation}
    \label{eq:Robasic}
    \mathrm{Ro} \equiv \frac{\mathrm{Rotation\,Timescale}}{\mathrm{Convective\,Timescale}},
\end{equation}
so that a system subject to significant Coriolis force possesses low Ro.  Convection that is relatively insensitive to rotation is characterized by a high value of Ro.  Once Ro exceeds some critical value, the convective Reynolds stress and meridional flow change such that a solar-like differential rotation is no longer sustainable \citep{Gilman1977,2007Icar..190..110A,kapyla2011,Gastine2013,2014MNRAS.438L..76G,Guerrero2013,Kapyla2014,Featherstone2015,FH16rot}.   

The Rossby number is typically defined in one of two ways (we defer precise definitions for both to \S\ref{sec:experiment}).  Much like the Reynolds number, it can be computed using the characteristic speed and length scale of the resultant flow. It can also be estimated a priori by system control parameters, effectively using the freefall time across the domain as a proxy for the convective timescale.  The latter formulation is referred to as the \textit{convective} Rossby number, which we denote using a subscript ``c,'' as \Roct.

In what is perhaps the most extensive examination of the topic to date, \citet{2014MNRAS.438L..76G} incorporated data from multiple rotating-convection studies and found that the transition between regimes corresponds to a unity value for \Roct~(i.e., when buoyancy and Coriolis forces are approximately equal).  This general behavior appears to be independent of other system properties, such as shell aspect ratio or thermal- and velocity- boundary conditions.  It also appears to be relatively insensitive to magnetism; MHD studies indicate only a slight shift and/or broadening of the \Roct=1 transition due to the presence of the Lorentz force \citep{Fan2014,Karak2015,Mabuchi2015,Simitev2015,Viviani2018,Warnecke2018,Viviani2021}.

\subsection{A Complementary View of the Transition}

That the transition point between solar and antisolar differential rotation depends on the relative strength of Coriolis and buoyancy forces is in many ways unsurprising.  Convection subject to strong rotational influence is characterized by organized and anisotropic transport of angular momentum owing to the development of columnar convective structures \citep{Zhang1992,Busse2002}.  As discussed in \citet{2007Icar..190..110A}, such correlated structures do not arise in the absence of strong rotational constraint, and the resulting convection tends to mix angular momentum throughout the shell, leading to the antisolar configuration.

Through this work, we further examine the link between convective structure and the solar/antisolar transition.  By considering a suite of numerical models with a range of convection-zone depths, we demonstrate that the point of transition occurs when the characteristic spatial scale of convection exceeds the depth of the convective layer.  At that point, the coherent columnar structures required to sustain a rapidly-rotating equator no longer manifest.  We find that this geometric criterion for the transition is consistent with, and indeed complementary to, the \Roct=1 criterion that has been found in previous studies.

 We provide a description of our numerical approach and suite of new numerical models developed for this study in \S\ref{sec:num}, followed by a presentation of the results in \S\ref{sec:results}.  In \S\ref{sec:interpretation} we present a force-balance analysis that illustrates the rough equivalence between our structural/geometric criterion and the previously observed \Roct$=1$ point of transition.

\section{The Numerical Experiment}\label{sec:num}
\subsection{Anelastic Formulation}\label{sec:formulation}
For this study, we choose to model nonmagnetic, rotating convection in spherical shells under the anelastic approximation \citep{1953QJRMS..79..224B,1969JAtS...26..448G,1981ApJS...45..335G}. This approach retains the effects of compressibility arising from background density and temperature stratification, while filtering out acoustic modes.  It is appropriate for the study of deep stellar and planetary interiors where perturbations about the thermodynamic background state are small and fluid motions are subsonic. 

All models presented here employ a thermodynamic background state satisfying the ideal gas law such that
\begin{equation}
    \label{eq:eos}
    \Pbar= \mathcal R \rhobar  \Tbar.
\end{equation}
Here, horizontal overlines indicate background-state quantities, $P$ is the pressure, $\rho$ is the density, $T$ is the temperature and $\mathcal R$ is the gas constant. Fluctuations about the background profile are indicated by the absence of an overline.  They may be related by linearizing Equation \ref{eq:eos}, yielding
\begin{equation}
\label{eq:linear_eos}
\frac{\rho}{{\rhobar}}= \frac{P}{\Pbar} - \frac{T}{\Tbar}= \frac{P}{\gamma \Pbar} - \frac{S}{c_p},
\end{equation}
where $S$ is the specific entropy, $c_p$ is the specific heat at constant pressure and $\gamma$ is the adiabatic index.

The anelastic continuity equation is given by
\begin{equation}
    \label{eq:continuity}
    \vnabla \cdot \left(\rhobar \vvec\right)=0,
\end{equation}
where ${\bm{v}}=\left(v_r,v_\theta,v_\phi\right)$ is the velocity vector in spherical coordinates.  Its evolution is described through the momentum equation
\begin{equation}
\label{eq:momentum}
\rhobar \left( \frac{D {\bm{v}}}{Dt}+2 {\Omega_0\bm{\hat{z}}} \times {\bm{v}}\right)=-\rhobar\, \vnabla\frac{P}{\rhobar}-\frac{\rhobar S}{c_p} {\bm{g}}   +\vnabla\cdot\boldsymbol{\mathcal{D}}, 
\end{equation}
where ${\bf g}$ is the gravitational acceleration, $\bm{z}$ is unit vector in the z-direction (parallel to the rotation axis), and $\Omega_0$ is the frame rotation rate.  The viscous stress tensor $\boldsymbol{\mathcal{D}}$ is given by:
\begin{equation}
  \label{eq:stress}
   \mathcal{D}_{ij} = 2\,\rhobar\,\nu\left[e_{ij}-\frac{1}{3}(\vnabla\cdot\bm{v})\delta_{ij}\right],
\end{equation}
where $\nu$ denotes the kinematic viscosity, $e_{ij}$ is the strain stress tensor, and $\delta_{ij}$ is the Kronecker delta.  The combined form of buoyancy and pressure appearing in Equation \ref{eq:momentum} is exact for adiabatically stratified background states such as those employed in our models.  It remains a reasonable approximation for background states that are weakly non-adiabatic as well \citep{Lantz1992,1995GApFD..79....1B}.

Finally, the evolution of $S$ is described by
\begin{equation}
\label{eq:thermal}
\begin{split}
  \rhobar\Tbar\frac{D S}{D t} =&
\vnabla\cdot\left[\rhobar\Tbar\kappa\,\vnabla S \right] +Q \\ &+  2\, \rhobar\, \nu \left[ e_{ij} e_{ij} -\frac{1}{3} \left(\vnabla \cdot \vvec\right)^2 \right],
\end{split}
\end{equation}
where $\kappa$ is the thermal diffusivity, and where $Q$ denotes any possible source or sink of internal energy, such as that which might arise through nuclear burning or radiative heating.

\subsection{Numerical Approach}\label{sec:approach}

We evolve the system of equations \ref{eq:eos}--\ref{eq:thermal} using the open-source Rayleigh convection code \citep{Rayleigh_2021}.  Rayleigh solves these equations in 3-D spherical geometry using a spectral transform approach based on that described in \cite{1984JCoPh..55..461G}.  System variables are represented radially using a truncated expansion of Chebyshev polynomials $T_n(r)$ extending up to maximum degree $n_\mathrm{max}$.  A truncated expansion in spherical harmonics $Y_\ell^m(\theta ,\phi)$, extending up to maximum Legendre degree $\ell_\mathrm{max}$,  is employed on spherical shells at each radius.  Both polynomial expansions are dealiased in radius such that
\begin{equation}
    \label{eq:dealias}
    n_\mathrm{max}+1 = \frac{2}{3}N_r\,\,\,\mathrm{and}\,\,\,\ell_\mathrm{max}+1 = \frac{2}{3}N_\theta,
\end{equation}
where $N_r$ and $N_\theta$ are the number of radial and latitudinal collocation points employed respectively.
Derivatives in radius and on spherical surfaces are calculated using the properties of these two basis sets respectively.
Time-integration is accomplished using a hybrid implicit/explicit  scheme with linear and nonlinear terms evolved using the Crank-Nicolson and the Adams-Bashforth methods respectively.  The solenoidal constraint on the mass flux described by Equation \ref{eq:continuity} is satisfied by decomposing the velocity field into streamfunctions such that
\begin{equation}
\label{eq:stream}
\rhobar\vvec= \vnabla \times \vnabla \times (W \bm{\hat{r}}) + \vnabla \times (Z\bm{\hat{r}}),
\end{equation}
where $\bm{\hat{r}}$ is the radial unit vector.  $W$ and $Z$ are the poloidal and toroidal streamfunctions respectively.

\subsection{Model Setup}\label{sec:experiment}

For this study we simulate a series of 18 model stellar convective zones. Each simulation is initiated with a polytropic thermodynamic background state using the prescription of \cite{2011Icar..216..120J}.  Following \citet{FH16norot}, we select a set of polytropic parameters that describe an adiabatically stratified background state that resembles the solar convection zone in many respects.  Specifically, we adopt a polytropic index $n$ of 1.5, an interior mass of 1.989$\times10^{33}$ g, a density variation $N_\rho$ of 3 density scaleheights spanning the convective shell, a density of 1.805$\times10^{-1}$ g cm$^{-3}$ at the inner boundary, and a value of 3.5$\times10^8$ erg K$^{-1}$ g$^{-1}$ for $c_p$.  As noted in \citet{FH16norot}, adopting this formulation in combination with solar-like values for the domain bounds yields a thermodynamic profile in good accord with that determined helioseismically for the Sun \citep[e.g.,][]{modelS}.

Through this study, we expand on the results of \citet{FH16rot} and \citet{Hindman20}, who employed a shell aspect ratio $\chi$ of 0.759, corresponding to inner and outer radius of $r_{\rm inner}=5 \times 10^{10} \, \rm cm$ and  $r_{\rm outer}=6.586 
\times 10^{10} \, \rm cm$.   The models presented in this study supplement those earlier datasets by varying the outer convection-zone radius, and thus the shell aspect ratio.  We examine one set of models possessing a thin convection zone, where $\chi=0.85$ and $r_{\rm outer}=5.88 \times 10^{10} \, \rm cm$.  We generate and analyze a second set of models as well, this time possessing thick convection zones with $\chi=0.38$ and $r_{\rm outer}=13.16 \times 10^{10} \, \rm cm$.

For each model, we adopt  impenetrable, stress-free
boundaries conditions.  At the lower boundary, we employ thermally-insulating boundary conditions ($\partial S / \partial r = 0)$ whereas we enforce a fixed-entropy condition at the upper boundary ($S=0$).  As in \citet{FH16norot}, each model possesses an internal source of heat $Q(r)$ such that
\begin{equation}
    \label{eq:heating}
    Q(r,\theta,\phi) = A\left(\Pbar(r)-\Pbar(r_{\mathrm{inner}})\right).
\end{equation}
The normalization constant $A$ is chosen so that
\begin{equation}
    \label{eq:norm}
    L_\star = \int_V Q(r,\theta,\phi) dV,
\end{equation}
where $L_\star$ is the stellar luminosity.  All simulations presented here are nonmagnetic.   {Following thermal and dynamical equilibration, all models were further evolved for at
least $2/3$ of a viscous diffusion timescale across the layer. In practice, the thin shell models were run for several tens of viscous diffusive times due to the reduced diffusion time across the thin shell.  Similarly, the models of \citet{FH16rot} and \citet{Hindman20}, whose data we incorporate into this study, were evolved for at least one viscous timescale following equilibration.}

\thintable

\thicktable

We vary the rotation rate, luminosity and diffusivities across our series of models which are equivalently described by three nondimensional numbers:  a Prandtl number (Pr), Ekman number (Ek), and a flux Rayleigh number (Ra$_\mathrm{F}$).  The Prandtl number,  which expresses the relative strength of viscous and thermal diffusion, is defined as
\begin{equation}
    \label{eq:Prandtl}
    \mathrm{Pr} = \frac{\nu}{\kappa}.
\end{equation}
We adopt a value of unity for Pr in all simulations in this study.   The Ekman number Ek, which expresses the ratio of the rotational and viscous timescales is given by 
\begin{equation}
    \label{eq:Ek}
    \mathrm{Ek}=\frac{\nu}{2\Omega L^2},    
\end{equation}
where $L$ is the shell depth.   The Rayleigh number expresses the strength of buoyancy relative to diffusive processes and is defined as
\begin{equation}
    \label{eq:RaF}
 \mathrm{Ra}_\mathrm{F}=\frac{\tilde{g} \tilde{F} L^4}{c_p \tilde{\rho} \tilde{T} \nu \kappa^2},
\end{equation}
where tildes indicate a volume-averaged value for the underlying variable.  We denote our Rayleigh number with a subscript ``F'' to indicate that the entropy scale is defined in terms of the fixed flux imposed through the system.  That flux, which convection must transport in response to the heating $Q$, is denoted by $F$.  The values of $\nu$, $\kappa$ and $c_p$  are taken to be constant functions of space in this study.  

An additional nondimensional number, the convective Rossby number Ro$_\mathrm{c}$, characterizes the relative strength of buoyancy and Coriolis forces.  It may be expressed in terms of the other three control parameters as
\begin{equation}
    \label{eq:Roc}
    {\rm Ro_c} \equiv \sqrt{\frac{\rm Ra_FE^2}{\rm Pr}},
\end{equation}
and it provides an a priori estimate of the degree to which rotation influences the convection.   Once the system has equilibrated, that rotational influence can be measured directly via the system-scale Rossby number Ro, namely
\begin{equation}
    \label{eq:Ro}
    {\rm Ro} = {\rm ReEk} = \frac{\tilde{U}}{2\Omega L},
\end{equation}
where $\tilde{U}$ is a characteristic velocity amplitude associated with the equilibrated system and Re=$\tilde{U}L/\nu$ is the system-scale Reynolds number.  For $\tilde{U}$, we adopt the rms convective velocity amplitude, removing the azimuthally-symmetric component and taking the rms mean over the full spherical shell.  A complete list of parameters for all models is provided in Tables 1 and 2.  { In addition, simulation checkpoints, system snapshots, and time-averaged outputs for each model may be accessed at}{\color{blue} \href{https://osf.io/j275z/wiki/Dataset\%20Listing/} {https://osf.io/j275z/wiki/Dataset\%20Listing/}}.

\section{Results}\label{sec:results}

\subsection{Solar, Antisolar, and Transitional Regimes}

\thinazavg

\thickazavg

\thermal

As our goal is to examine the solar/antisolar transition, we have classified all models described in Tables 1 and 2 as solar (``S''), antisolar (``A'') or transitional (``T'').  In addition to distinct differential rotation profiles, the solar and antisolar states represent two basins of attraction that possess distinct structuring of thermal gradients and meridional circulation as well.  We define a solar-like state as possessing three characteristics:
\begin{enumerate}
    \item An equator of prograde rotation, and polar regions with retrograde rotation in the rotating frame.
    \item Polar regions that are warm relative to the equatorial region.
    \item Meridional circulations that possess multiple cells in depth.
\end{enumerate}
The antisolar state is defined as one that possesses:
\begin{enumerate}
    \item An equator of retrograde rotation and polar regions that rotate prograde in the rotating frame.
    \item Polar regions that are cool relative to the warmer equatorial regions.
    \item Meridional circulations that are primarily monocellular within a hemisphere.
\end{enumerate}
The relationship between the meridional flow and the differential rotation profiles stems from the fact that meridional transport of angular momentum must balance the convective Reynolds stress in a steady-state system \citep{Miesch2011,Featherstone2015}.   Thermal profiles in the low-Ro, solar-like regime are established due to the fact that such systems tend  to be in thermal-wind balance and possess columnar convection that transports heat more efficiently in the polar regions \citep[e.g.,][]{Brun2002,Matilsky2020}.  In antisolar systems, convection is more efficient in the equatorial regions, resulting in a warmer equator \citep[e.g.,][]{Featherstone2015} .

Examples of these different states for the two different shell geometries are illustrated in Figures \ref{fig:thin_azavg} and \ref{fig:thick_azavg}. Antisolar states are illustrated in the left column, solar-like states on the right, and example transitional states are shown in the central column.  Intermediate regimes between the solar and antisolar classes of behavior have previously been found in studies examining the transition \citep[e.g.,][]{Gilman1977,Glatzmaier1982,2007Icar..190..110A,Gastine2013}.  For this study, we define `transitional' to mean a system whose mean thermal and/or flow profiles deviate from the definitions provided above. % {It is important to remark that all these models have been evolved for a minimum of 2/3 viscous diffusion times following the onset of a statistically-steady state. Therefore, the term 'transitional' does not indicate that these models will eventually reach another state, but it indicates a steady-state itself which combines characteristics of both antisolar and solar states.}

\citet{Featherstone2015} identified one example of such an intermediate regime, characterized by solar-like differential rotation occurring in the presence of a monocellular meridional flow.  Case S3 (Figure \ref{fig:thin_azavg}, central column) provides another example.  The meridional flow and latitudinal entropy gradients satisfy the antisolar definition.  The differential rotation, however, is antisolar in the upper convection zone, but solar-like in the lower convection zone.

In our thick-shell models, transitional behavior seems to appear in the thermal profiles.  Case L3 (Figure \ref{fig:thick_azavg}) serves as one example of this behavior.  While its differential rotation and meridional flow are clearly antisolar in nature,  its specific entropy profile is opposite of that expected for the antisolar state and is instead akin to the solar state. 

A close inspection of case L3 reveals a thin boundary layer possessing a roughly antisolar-like specific entropy profile.  As \Roct~is increased further, this boundary layer gradually extends throughout the domain (see Figure \ref{fig:thermal}).  Eventually, as in case L1, the bulk of the domain possesses cool poles and a warm equatorial region, though the region interior to the tangent cylinder remains warm in the deep convection zone.  While still possessing some semblance of transitional behavior, we choose to define L1 as ``antisolar''.   We note that a similar range in \Roct~(and resulting Ro) is covered by the sets of models L1--L4 and S1--S4 that span the transition, but the thermal behavior discussed above is only observed in the thick-shell models.

\subsection{Connecting Mean Flows and Convective Structure}

We find that the transition between the solar and antisolar regimes can be described in terms of the convective structure as characterized through its power spectrum.   Following \citep{FH16rot}, we consider the power spectrum associated with horizontal flows, subtracting the contribution from axisymmetric differential rotation and meridional circulation.  Namely, we consider convective power $P_\ell$ defined as
\begin{equation}
    \label{eq:convective_power}
     P_\ell(r)=\sum_{\substack{m=-\ell \\ m\ne0}}^{\ell} \left(|u_{\ell,\theta}^{m}(r)|^2 + |u_{\ell,\phi}^{m}(r)|^2\right).
\end{equation}
Here the $u_{\ell\,,\,j}^m(r)$ are the complex coefficients resulting from the expansion of velocity field components into $Y_\ell^m$'s on spherical surfaces of radius $r$.

\thinflows

\thickflows

The convective power for our thin- and thick-shell models is illustrated in Figures \ref{fig:thin} and \ref{fig:thick}, respectively.   Power  has been sampled near the upper boundary and time-averaged over at least one-third of the viscous diffusion time-scale in all cases.  Representative flow patterns from each series are also illustrated in the top panels of these figures.   With the exception of cases S7 and S8, all spectra possess a single, broad peak characterized by a central value $\ell_\mathrm{peak}$.  We measure the value of $\ell_\mathrm{peak}$ by fitting the power-spectrum peak with a fourth-order polynomial in $\ell$.  Note that cases S7 and S8 possess multiple secondary peaks, a common feature for systems near convective onset \citep{Hindman20}. 

In both series, as \Roct~is decreased, peak spectral power occurs at higher values of spherical harmonic degree $\ell$.  This trend was also observed in \citet{FH16rot}.  It arises from the tendency of rotating convection to organize into columnar structures with increasingly smaller cross-sectional diameter as the Rossby number is decreased  \citep[e.g.,][]{Busse2002}.  

In Figure \ref{fig:transition}, we illustrate the relationship between convective spatial scale and the solar/antisolar transition. There, we present our simulation results in terms of the control parameter \Roct~  and the resulting convective scale $\lpeak$.  We also include data from \citet{FH16rot} and \citet{Hindman20}.  Those models possessed an identical polytropic background state to the models presented here, but with an intermediate convection zone thickness of $\chi=0.76$.  

 {
The models of \citet{Hindman20} were originally classified based on their convective, rather than mean-flow, structure.  We have found that many of the more laminar models, classified as ``equatorial columns'' and ``modulated convection,'' largely fall into either the solar or transitional regimes.   In generating Figure \ref{fig:transition}, we reclassified those results according to the criteria described above, omitting data from eight models: model numbers 6--9,17,18,29,30.  The steady-state mean flows in those systems (all near convective onset) were difficult to characterize with certainty due to strong, but slowly-varying hemispheric asymmetries in their flow and thermal profiles.
}

\transitionfig

As in \citet{2014MNRAS.438L..76G}, we find that the transition between solar-like (red symbols) and antisolar differential rotation (blue symbols) occurs when \Roct~is roughly unity.  We also find that the transition is equally well-characterized by the point at which the dominant convective spatial scale is roughly equal to the shell depth.  Namely, the transition occurs when the characteristic convective wavelength and the shell depth differ by half a wavelength, such that
\begin{equation}
    \label{eq:constraint}
    \ell_\mathrm{peak}=\ell_\mathrm{shell}/2,    
\end{equation}
 where
\begin{equation}
    \label{eq:lshell}
    \ell_\mathrm{shell}=2\pi r_\mathrm{outer}/L,
\end{equation}
and where $L$ is the shell depth.  As discussed, for those models in the low- and intermediate Ro- regimes, the spatial scale $\ell_\mathrm{peak}$ is associated with convective columns.   The  { constraint in equation}  \ref{eq:constraint} thus indicates that the transition occurs when the shell depth and the characteristic convective column diameter, which is half the convective wavelength, are equal.  This is the central result of this paper, and to our knowledge, it has not been noted or discussed before.  In the following section, we discuss why this geometric constraint is to be expected and explore how it relates to the equivalent, and more widely-reported Ro$_\mathrm{c}$=unity criterion for the transition.

\newcommand{\geff}{\Hat{g}}

\section{Interpretation}\label{sec:interpretation}

\FsFz

\composite

The link between shell depth, convective spatial scale, and the solar/antisolar transition can be understood by considering the convective transport of angular momentum in the high- and low-Ro regimes.  The specific angular momentum about the z-axis, $\mathcal{L}$, is given by
\begin{equation}
\label{eq:Lspecific}
\mathcal{L}=\lambda^2\left( \Omega_0+\frac{\langle v_\phi \rangle}{\lambda} \right),
\end{equation}
where $\lambda=r\mathrm{sin}\theta$ is the cylindrical radius, and $\Omega_0$ is the frame rotation rate. 
Neglecting viscous and Lorentz torques, the time-evolution of zonally-averaged $\mathcal{L}$ is described by
\begin{equation}
\label{eq:amom}
\frac{\partial {\bf \mathcal{L}}}{\partial t}=-\vnabla\cdot\boldsymbol{F}_{RS} - \rhobar\langle \vvec \rangle\cdot\vnabla {\bf \mathcal{L}}.
\end{equation}
Here, $\boldsymbol{F}_{RS}$ is the convective Reynolds stress, defined as
\begin{equation}
\label{eq:Reynolds}
\rhobar\lambda\langle \vvec' v_\phi' \rangle
\end{equation}
where angular brackets indicate a zonal (azimuthal) average, and where primed quantities indicate fluctuations about the zonal average.

In a steady state, the two terms on the RHS of Equation \ref{eq:amom} balance, providing a direct link between the structure of meridional flow $\langle v_r, v_\theta \rangle$ and the convective flow structure \citep[see e.g.,][]{Miesch2011,Featherstone2015}.  In principle, the steady-state balance provides no information concerning the resulting, equilibrated differential rotation profile.  Equation \ref{eq:amom} does, however, illustrate the centrality of convective structure and amplitude, through the Reynolds stress it drives, to the redistribution of angular momentum.

The relation between convective structure and Reynolds stress to the transition is discussed in detail in \citet{2007Icar..190..110A}, who note an important difference between angular momentum transport in high-Ro and low-Ro convection.   In the high-Ro regime, the Reynolds stress is  { predominantly directed radially inward, and convective flows work to mix angular momentum throughout the domain.  This mixing is further enhanced by the strong meridional circulations driven by that convection \citep[e.g.,][]{Gilman1977,Featherstone2015}.}  

A system in solid-body rotation possesses most of its angular momentum in the equatorial regions, and so a uniform redistribution of that angular momentum requires a spin-up at the poles and slow-down at the equator.  The result is an antisolar differential rotation.  In practice, a perfectly-mixed state is never fully achieved in simulations, owing to the countervailing effects of meridional and viscous transport \citep[e.g.,][]{Featherstone2015}.

In the low-Ro regime, convection is markedly anisotropic in nature \citep[e.g.,][]{Zhang1992,Busse2002}.  In such systems, convection organizes into columnar rolls that exhibit a preferred ``tilt'' in the eastward (positive $\phi$) direction.  This tilting results from the tendency of columnar upflows and downflows to conserve potential vorticity as they approach or descend from the spherical boundary \citep[see also the discussion in][]{2007Icar..190..110A}.  These tilted flow structures establish a positive correlation between flows moving outward (inward) from the rotation axis and flows moving in the positive (negative) $\phi-$direction.  A net angular-momentum transport away from the rotation axis results, and so low-Ro convection tends to speed up the equator until counterbalanced by meridional or viscous transport.  

 {
The Reynolds stresses arising from a selection of our thick-shell models are shown in Figure \ref{fig:FsFz}.  There, we decompose the longitudinally-averaged convective Reynolds stress into cylindrical coordinates ($s$,$z$), where the radial direction ${\bf \hat{s}}$ is perpendicular to the rotation axis, and where ${\bf \hat{z}}$ is parallel.   In the low-Ro models L8 and L10, the Reynolds stress is predominantly orthogonal to the rotation axis and directed cylindrically outward.  While transport in the perpendicular ${\bf z}$-direction is present, its amplitude is weaker by roughly one order of magnitude.}   

 {
This situation contrasts with that arising in the transitional and high-Ro regimes represented by models L3 and L1, respectively.  There, convection preferentially transports angular momentum inward, both toward the rotation axis and toward the equatorial plane, with little difference in amplitude between the ${s}$- and ${z}$-transport.  Only in the low-Ro systems is a strong anisotropy in angular momentum transport realized. 
}

 {We further illustrate this behavior schematically, alongside instantaneous snapshots of the fluid,} from a selection of our thin-shell simulations in Figure \ref{fig:composite} (see also Figure 3 of \citet{Simitev2015}).  There we show three models, with \Roct~values ranging from 0.346 to unity (the point of the transition).  At low \Roct, convective flows (as visualized in equatorial cuts of z-vorticity, $\omega_z$) exhibit noticeable prograde tilting such that flow patterns near the outer boundary are shifted eastward relative to the lower boundary.   As a result,  {the radial component of the} Reynolds stress in the low-\Roct-regime (Figure \ref{fig:composite}, center row), is primarily positive.  The coherence of this correlation gradually diminishes as \Roct~increases and convective cells approach an aspect ratio of unity at \Roct=1, beyond which point most spatial coherence is lost.  

When this key element of anistropic angular momentum transport is lost, there is no additional source of anistropy available to establish a rapidly-rotating equator.  We suggest that this loss of correlation can be understood from a geometric standpoint.  For a columnar convective cell to exhibit tilting, its size in the $\phi$-dimension must be less than its extent in depth.  Once its horizontal extent is equivalent to the shell depth, and the cell attains an aspect ratio of one, this situation is no longer possible.  At that point, for the columnar structure to exhibit tilting, its extent in depth would need to exceed the depth of the convective layer.  In the next subsections we explore the relationship between this geometric view of the transition and the \Roct=unity criterion that has been identified previously.

\subsection{Characteristic Timescales}
First, we consider some relevant timescales and their relationship to our nondimensional control parameters.  The viscous and thermal diffusion timescales across the layer, $\tau_\nu$ and $\tau_\kappa$ respectively, are given by
\begin{equation}
\label{eq:taunu}
\tau_\nu=\frac{L^2}{\nu}
\end{equation}
and
\begin{equation}
\label{eq:taukappa}
\tau_\kappa=\frac{L^2}{\kappa},
\end{equation}
where, again, $L$ is shell depth.  The Coriolis timescale $\tau_\Omega$ is given by
\begin{equation}
\label{eq:tauomega}
\tau_\Omega = \frac{1}{2\Omega},
\end{equation}
and the timescale $\tau_{ff}$ for a fluid parcel to freely fall across the domain depth given by
\begin{equation}
\label{eq:freefall}
\tau_{ff} = \sqrt{L/\geff}.
\end{equation}
Here, $\geff$ is the effective gravitational acceleration due to buoyancy, namely
\begin{equation}
    \label{eq:geff1}
    \geff = g\frac{\rho'}{\rhobar},    
\end{equation}
where $g$ is the gravitational acceleration,  $\rho'$ is a characteristic density perturbation and $\rhobar$ is the background density. 

Using $\geff$, we write the Rayleigh number Ra in general form as
\begin{equation}
\label{eq:ratime}
\mathrm{Ra}=\frac{\geff L^3}{\nu\kappa}=\frac{\tau_\nu\tau_\kappa}{\tau_{ff}^2}= \mathrm{Pr}\left(\frac{\tau_\nu}{\tau_{ff}} \right)^2.
\end{equation}

This form of Ra can be used regardless of the particular expression adopted for the relative density perturbation.  For the anelastic, fixed-flux models presented here, we have chosen to write the relative density perturbation as
\begin{equation}
    \label{eq:geff1}
 \frac{\rho'}{\rhobar}=\frac{\tilde{F} L}{c_p \tilde{\rho} \tilde{T} \kappa}.
\end{equation}
This yields the Expression \ref{eq:RaF} for $\mathrm{Ra}_\mathrm{F}$, where the subscript $F$ is use to indicate that a flux-based scaling was adopted for the density perturbations.  In a Boussinesq fluid, with fixed temperature contrast $\Delta T$ used as the relevant temperature scale, the density scale is typically chosen such that
\begin{equation}
\label{eq:geff2}
\frac{\rho'}{\rhobar} =\alpha\Delta T,
\end{equation}
where $\alpha$ is the coefficient of thermal expansion.  That prescription recovers the canonical expression
\begin{equation}
    \label{eq:Ra}
    \mathrm{Ra}=\frac{g\alpha \Delta T L^3}{\nu\kappa}.
\end{equation}

We can similarly recast Ek, and \Roct~in terms of characteristic timescales.  We have that
\begin{equation}
\label{eq:ektime}
\mathrm{Ek}=\frac{\nu}{2\Omega\,L^2}= \frac{\tau_\Omega}{\tau_\nu},
\end{equation}
and
\begin{equation}
\label{eq:Roctime}
\mathrm{Ro_c} = \sqrt{\mathrm{Ra}\,\mathrm{Ek}^2/\mathrm{Pr}} = \frac{\tau_\Omega}{\tau_{ff}},
\end{equation} 
which will be useful in the analysis that follows.

\subsection{Vorticity Dynamics}

%As shown in Figure \ref{f5}, the dominant horizontal length scales in our simulations relate to the system Rossby number Ro similarly to those of {\color{red} Featherstone \& Hindman 2016}.  Namely, the relation between $\ell_\mathrm{peak}$ reflects a balance between inertial and Coriolis forces. 
\scaling

Using Equation \ref{eq:Roctime} in combination with the vorticity equation, we now seek to relate \Roct~to the convective spatial scale and the layer depth.   We consider a fluid in the so-called CIA balance wherein viscous effects can be neglected, and the dominant force balance is struck between buoyancy, inertial, and Coriolis forces.  For a system in such a balance, the characteristic horizontal wavenumber of convection scales in proportion to $\mathrm{Ro}^{-1/2}$ \citep[see e.g.,][]{Ingersoll1982,FH16rot,Aurnou2020}.  

This balance is realized in many of the models considered in this paper, as illustrated in Figure \ref{fig:scaling}$a$. There, we plot the variation of $\ell_\mathrm{peak}$ for all models considered in this study.  { The logarithm  of reduced Rayleigh number $\mathrm{Ra}^*=\mathrm{Ra}_\mathrm{F}\,\mathrm{Ek}^{4/3}$ is indicated graphically by the symbol shading such that symbols with lighter shading have lower values of $\mathrm{Ra}^*$.  As this parameter is increased, the convective length scale tends toward the $\mathrm{Ro}^{-1/2}$ curve, a fact also noted in \citet{Hindman20}.}

 {We note that this trend in the scaling of convective length scale is not completely unambiguous. As illustrated in Figure \ref{fig:scaling}$b$, all solar-like models, regardless of $\mathrm{Ra}^*$, possess convective length scales that vary in accord with the $\mathrm{Ek}^{-1/3}$ scaling associated with convective onset \citep{Chandra1953}.   We find that only those models with $\mathrm{Ra}^*\ge10$ exhibit a dual scaling. 
}    

 {
Motivated by the fact that many of our models exhibit behavior consistent with CIA balance, we now examine the transition from an inviscid viewpoint.}  Neglecting viscous effects and considering the $z$-component of the vorticity equation, we have that
\begin{equation}
    \label{eq:vortz1}
    2\Omega\frac{\partial v_z}{\partial z} \sim \vvec \cdot \vnabla \omega_z.
\end{equation}

Considering a columnar convective cell, we take the axial ($z$) dimension of the column to be the shell depth $L$.  Denoting the column diameter (i.e., the lengthscale perpendicular to the rotation axis) by $D$, we have
\begin{equation}
    \label{eqvortz1p5}
    2\Omega\frac{\tilde{v}}{L}\sim\frac{\tilde{v}^2}{D^2},
\end{equation}
which reduces to
\begin{equation}
\label{eq:vortz2}
D^2 \sim \frac{\tilde{v}\,L}{2\Omega},
\end{equation}
for some characteristic velocity amplitude $\tilde{v}$.   By considering the solar/antisolar transition, we are implicitly considering a point of transition between slowly-rotating and rapidly-rotating regimes of convection.  As discussed in \citet{Aurnou2020}, there are two choices for $\tilde{v}$ corresponding to these two limits.   In the slowly-rotating limit, the flow speed is well-approximated via a freefall scaling such that
\begin{equation}
\label{eq:veff2}
\tilde{v_{ff}}\sim\frac{L}{\tau_{ff}}.
\end{equation}
Alternatively, in the rapidly-rotating limit, a thermal-wind scaling for $\tilde{v}$ is a more appropriate choice. Namely,
\begin{equation}
    \label{eq:vtwb}
    \tilde{v}_{TW}\sim\geff\tau_\Omega = \left(\frac{L}{\tau_{ff}}\right)\left(\frac{\tau_\Omega}{\tau_{ff}}\right),
\end{equation}
which differs from the free-fall estimate by a factor of \Roct.  Proceeding with the rapidly-rotating choice, the constraint that a convective column has diameter equal to the layer depth becomes
\begin{equation}
\label{eq:rc2}
L^2 = D^2 \sim \left(\frac{L}{2\Omega}\right)\left(\frac{L\tau_\Omega}{\tau_{ff}^2} \right)=L^2\left(\frac{\tau_\Omega}{\tau_{ff}}\right)^2
\end{equation}
which holds when:
\begin{equation}
\label{eq:equivalence}
\frac{\tau_\Omega}{\tau_{ff}} = \mathrm{Ro_c} = 1.
\end{equation}
A similar result is obtained when adopting a free-fall scaling for $\tilde{v}$, but the \Roct-factor arising in Equation \ref{eq:rc2} is no longer squared.

The estimate of Equation \ref{eq:equivalence} neglects prefactors of order unity but nevertheless provides a link between the complementary views of the solar/antisolar transition illustrated through Figure \ref{fig:transition}.  It is interesting to note that while Figure \ref{fig:transition} indicates a clear dependence on the shell depth,  \Roct=1 serves as an apparently shell-depth-insensitive marker of the transition.  This fact was also noted by \citet{2014MNRAS.438L..76G} who considered results from studies that employed different shell depths, as well as different boundary conditions and fluid approximations.   The reason for this can be seen in the expression for \Roct~in terms of timescales provided  by Equation \ref{eq:Roctime}.  \Roct~implicitly contains information regarding the shell depth and the fluid approximation employed via $\tau_{ff}$.

%\subsection{Non-dimensional numbers}

%In this Section we explain why our results do not contradict with the 
%previous results of Gastine 2014. These authors stated that the 
%transition from solar to anti-solar regimes occurs when $Ro_c \sim 1$, 
%regardless of the convective shell depth. 
%This argument is line with our results, as we found
%that, near the transition, $Ro_c\sim 2.8$ for the thin shell models and $Ro_c\sim 1.4$
%for our thick shell models.

\section{Summary and Perspectives}

Through this work, we have examined the relationship between convective spatial scale, convection-zone depth, and the solar/antisolar transition.  Our numerical results suggest that the transition occurs when the dominant convective wavelength is a factor of two larger than that associated with the convection zone depth.  Motivated by the anistropic nature of low-Ro convective Reynolds stress, we suggest the transition can be understood in terms of the flow correlations established by columnar convective structures that naturally arise in rapidly-rotating regimes.  As the moderate-Ro, slowly-rotating regime is approached, the characteristic columnar size exceeds the depth of the convective layer, and the structure can no longer manifest.  Transport of angular momentum becomes increasingly isotropic as a result, leading to an antisolar differential rotation.  This criterion of the transition is complementary to the previously identified \Roct=unity criterion.  When considered in terms of the CIA force balance, the spatial-scale criterion can be used to arrive at the \Roct=unity criterion.

%One effect not incorporated into this study was that of variable density stratification across the domain.  The latter was explicitly examined by \citet{kapyla2011}, who found that the transition occurred at similar values of Ro, regardless of the degree of stratification.  As a result, we see reason why the Ro-phenomenon explored here would change for models that were more or less stratified.

 {
All models presented in this study were run with a Prandtl number of unity.  While we see no obvious reason that this parameter would fundamentally impact the geometric picture presented here, we could but speculate on the effects of that control parameter based on this survey of models.  We note that the study of \citet{2014MNRAS.438L..76G}, incorporated results from models possessing Pr with values less than unity and found that they too transitioned at \Roct=1.  We also point the reader to the recent work of \citet{Kapyla2022}.  That study examined a range of Pr values, and it found that the transition point shifted for Pr$> 1$ models that incorporated magnetism. The inclusion of magnetism, which can alter the flow structure, makes a straightforward comparison against the results presented here difficult, but it is interesting to ask how its effects may impact the picture presented here.  
}

Magnetic effects are of particular interest due to the possible link between the solar/antisolar transition and observations of Ro-dependent changes in magnetic topology and activity in low-mass stars \citep[e.g.,][]{Brandenburg2018,Lehtinen2021}.  Studies incorporating magnetism have largely focused on dynamo behavior and the \Roct-point of the transition, but changes in the convective spectrum, induced by magnetism, can be seen in the results of \citet{Simitev2015}.   { This effect becomes particularly pronounced when the characteristic spatial scale of magnetism is small with respect to that of realized by the convection \citep{Hotta2021,Hotta2022}}.  In those results, the amplitude of convective power is diminished in the presence of magnetism, and the dominant spatial scale of the convection shifts to higher-order wavenumbers.  It thus seems likely that including magnetism leads to a decoupling between the value of Ro from \Roct.  This effect was observed in the magnetic study of \citet{Mabuchi2015}.  While that work did not consider the convective length scale explicitly, it demonstrated that the system-scale Ro associated with the transition remained unchanged in the presence of magnetism.  The transitional value of \Roct, however, was modified. 

We conclude by noting that our results bear some relationship to the so-called ``convective conundrum,'' a term given to a set of problems related to disagreements between models and observations of the Sun's convective flow structure and speed \citep[e.g.,][]{Omara2016}.  One aspect of this problem is that models driven toward ostensibly more solar-like conditions (i.e., higher Rayleigh and lower Ekman numbers), tend to develop an antisolar differential rotation.   This appears to be linked to an excess in convective power at large scales in many models.

Solar convection-zone models with differential rotation approximating that of the the Sun tend to possess substantial convective power on large spatial scales (typically only somewhat smaller than the convection-zone depth; $\ell\approx20$).  This is not born out through observations, however.  In the solar photosphere, convective power instead peaks on the much smaller spatial scale of supergranulation \citep[$\ell\sim100$; e.g.,][]{Hart1956,Leighton1962,Rincon2018}.  Larger-scale photospheric flows are much weaker and tend to be dominated by inertial waves \citep{Hathaway2021,Gizon2021}.  Helioseismic analyses of convective flows at depth remain inconclusive; some indicate substantial large-scale convective power, and others a lack thereof  \citep{Hanasoge2012,Greer2015,Proxauf2021}.  

Several solutions to the conundrum have been proposed for the apparent lack of large-scale power, including rotational influence on the convection \citep{FH16rot,Vasil2021}, a convection zone that is weakly subcritical \citep{Brandenburg2016}, and Lorentz torques arising from small-scale magnetism \citep{Hotta2021}.  Our results provide a new, though not particularly illuminating, constraint on the solar convective structure.  Namely, given the nature of the Sun's differential rotation profile along with the fact that it does not appear to be transitional in nature, we do not expect to see substantial convective power on scales commensurate with the convection-zone depth. Improved observations of the deep solar meridional flow, novel experimental studies \citep[e.g.,][]{koulakis2018}, and better constraints on solar convective structure, such as might be achieved via observations of the Sun's polar regions, may provide additional insight into these topics.

%\appendix

\begin{acknowledgments}

This work was supported by NASA grants  (HGC) 80NSSC17K0008 and (LWS) 80NSSC20K0193 and the University of Colorado Boulder. Resources supporting this work were provided by the NASA High-End Computing (HEC) Program through the NASA Advanced Supercomputing (NAS) Division at Ames Research Center.  We thank the anonymous referee for several comments that improved the presentation of these results.  N. Featherstone would like to thank Mark Miesch, Keith Julien, Jonathon Aurnou, Bradley Hindman and Michael Calkins for many useful discussions over the years on this and related topics that helped to shape the perspectives presented in this work. 

\end{acknowledgments}

\bibliographystyle{aasjournal}
\bibliography{sample631,MHD_DR,Hydro_DR}


\end{document}